\documentclass[
twocolumn,
english,
aps,
prx,
longbibliography,
superscriptaddress,
amsmath,
amssymb,
floatfix,
]{revtex4-2}

\usepackage{amsthm}
\usepackage{amsfonts}
\usepackage{siunitx}
\usepackage{amsmath}
\usepackage{amssymb}
\usepackage{graphicx}
\usepackage{verbatim}
\usepackage[colorlinks]{hyperref}
\usepackage{tikz}
\usepackage{pgfplots}
\usepackage{adjustbox}
\usepackage{braket}
\usepackage{xcolor}
\usepackage{physics}
\usepackage{amssymb} 
\usepackage{graphicx}
\usepackage{dcolumn}
\usepackage{bm}
\usepackage{mathtools}
\usepackage{hyperref}
\usepackage{mathrsfs}
\usepackage{dashrule}
\usepackage{caption}
\usepackage{subcaption}
\usepackage{pifont}
\usepackage[version=4,arrows=pgf-filled,
textfontname=sffamily,
mathfontname=mathsf]{mhchem}

\newcommand{\Nqb}{N_{\text{qb}}}
\newcommand{\Ndw}{N_{\text{dw}}}
\newcommand{\ndw}{n_{\text{dw}}}
\newcommand{\Tm}{T_{\text{machine}}}
\newcommand{\Te}{t_{\text{eff}}}
\newcommand{\Teb}{\bar{t}_{\text{eff}}}

\newcommand{\Tp}{T_{\text{phys}}}
\newcommand{\Jp}{J_{\text{phys}}}
\newcommand{\Jen}{j_{\text{enc}}}

\usepackage[font=small,labelfont=bf,
   justification=justified,
   format=plain]{caption} 
   
\definecolor{linkcolor}{RGB}{0,83,166}
\hypersetup{
  colorlinks = true,
  allcolors = {linkcolor}
}
\def\thefootnote{\textsuperscript{\ding{168}}}\footnotetext{These authors contributed equally to this work.}

\begin{document}

\title{Classical Thermometry of Quantum Annealers}

\author{George Grattan\thefootnote{}}
\affiliation{{\em Q-M.A.F.I.A.}, Los Alamos National Laboratory, USA}
\affiliation{Analytics Division, Information Systems \& Modeling, Los Alamos National Laboratory}

\author{Pratik Sathe\thefootnote{}}
\email{psathe@dwavesys.com}
\affiliation{{\em Q-M.A.F.I.A.}, Los Alamos National Laboratory, USA}
\affiliation{Theoretical Division, Quantum \& Condensed Matter Physics, Los Alamos National Laboratory, USA}
\affiliation{Information Science and Technology Institute, Los Alamos National Laboratory, USA}
\affiliation{D-Wave Quantum Inc., Burnaby BCV5G 4M9, British Columbia, Canada}

\author{Cristiano Nisoli}
\email{cristiano@lanl.gov}
\affiliation{{\em Q-M.A.F.I.A.}, Los Alamos National Laboratory, USA}
\affiliation{Theoretical Division, Quantum \& Condensed Matter Physics, Los Alamos National Laboratory, USA}
\affiliation{Information Science and Technology Institute, Los Alamos National Laboratory, USA}
\affiliation{Center for Nonlinear Studies, Los Alamos National Laboratory, USA}

\date{\today}

\begin{abstract}
Quantum annealers are emerging as programmable, dynamical experimental platforms for probing strongly correlated spin systems. Yet key thermal assumptions, chiefly a Gibbs-distributed output ensemble, remain unverified in the large-scale regime. Here, we experimentally and quantitatively assess Gibbs sampling fidelity across system sizes spanning over three orders of magnitude. We explore a wide parameter space of coupling strengths, system sizes, annealing times, and D-wave hardware architectures. We find that the naively assumed scaling law for the effective temperature requires a non-negligible, coupling-independent offset that is robust across machines and parameter regimes, quantifying residual non-thermal effects that still conform to an effective Gibbs description. These non-idealities are further reflected in a systematic discrepancy between the physical temperature inferred from the sampled ensemble and the nominal cryogenic temperature of the device. Our results systematically assess the viability of quantum annealers as experimental platforms for probing classical thermodynamics, correct previous assumptions, and provide a physically grounded thermometry framework to benchmark these machines for future thermodynamic experiments.

\end{abstract}

\maketitle
In recent years, a broad range of strongly-correlated spin systems have found experimental realization on systems of coupled qubits, such as superconducting quantum annealers ~\cite{KairysSimulating2020,NarasimhanSimulating2024,lopez2024quantum,KingQubit2021,Lopez-BezanillaKagome2023,KingObservation2018}, neutral atoms platforms~\cite{EbadiQuantum2021,SchollQuantum2021}, trapped ions~\cite{PorrasEffective2004,BlattQuantum2012}, superconducting circuits~\cite{HouckOnchip2012}, and photonic materials~\cite{CarusottoPhotonic2020}. These machines have been employed as highly controllable ``Meccano sets'' for constructing and characterizing new quantum systems at the level of individual degrees of freedom, along lines initially proposed by R. Feynman~\cite{feynman2018simulating}. In particular, superconducting Quantum Annealers (QA) have led to direct vistas on exotic phenomenology of large scale systems, ranging from geometric frustration~\cite{KairysSimulating2020,NarasimhanSimulating2024,lopez2024quantum}, emergent monopoles~\cite{KingQubit2021,Lopez-BezanillaKagome2023}, topological phases~\cite{KingObservation2018},  fluctuation-induced correlations~\cite{lopez2024quantum}, and Kibble-Zurek mechanism~\cite{KingCoherent2022,KingQuantum2023a}. 

Although cloud users generally lack control over the device’s physical temperature, previous explorations have assumed an \emph{effective temperature}, controlled by the annealing parameters and even that the annealed ensemble is Gibbs-distributed~\cite{lopez2024quantum,KingQubit2021,Lopez-BezanillaKagome2023,KingObservation2018,AminSearching2015,SatheClassical2025,TezaFinitetemperature2025}.
Works have assessed the quality of these devices as Gibbs samplers~\cite{matsuda2009quantum,IzquierdoTesting2021, MarshallThermalization2017, LokhovOptimal2018,NelsonHighQuality2022, VuffrayProgrammable2022} and estimated the sampling's effective temperature~\cite{BenedettiEstimation2016a, RaymondGlobal2016}, but primarily for small systems.

\begin{figure*}[t!]
    \includegraphics[width=0.95\textwidth]{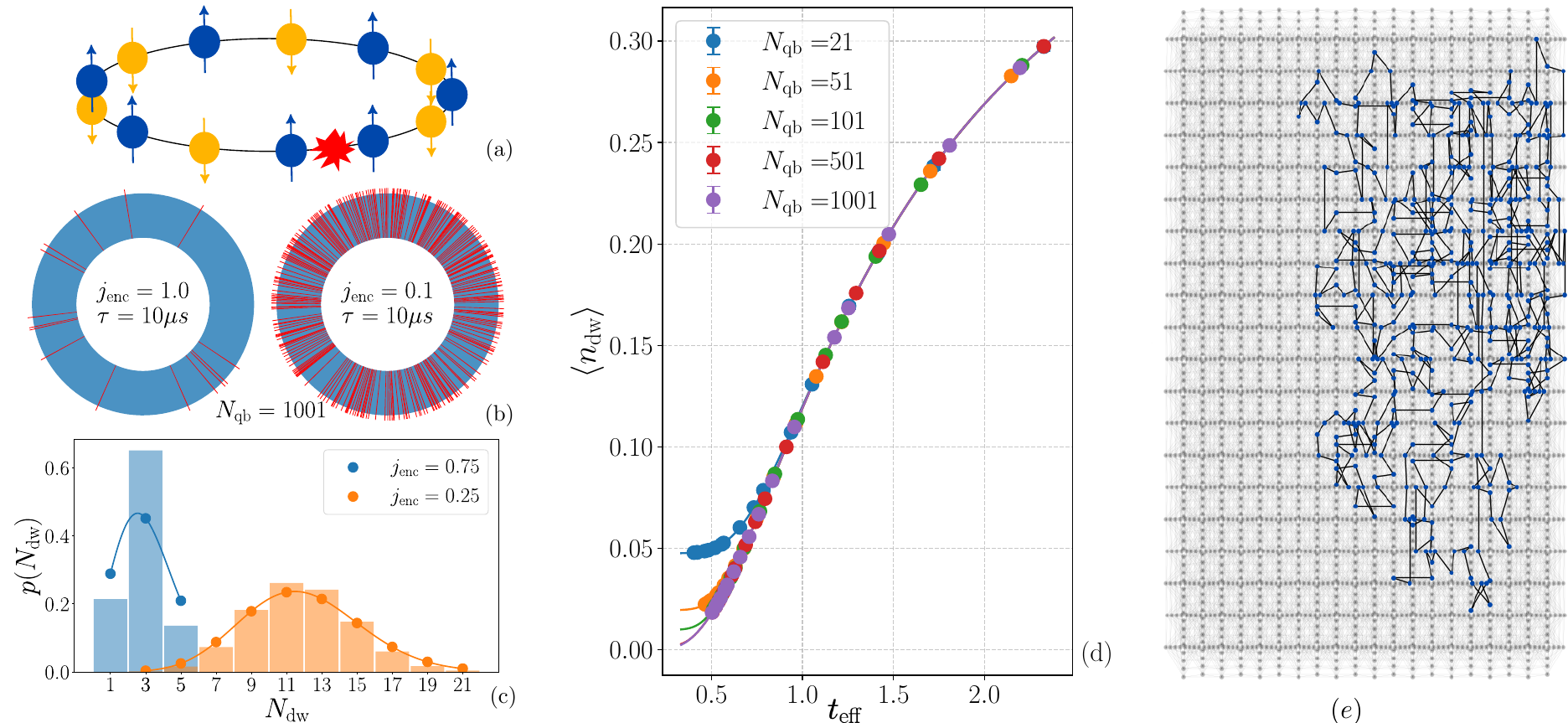}
    \caption{{\bf Model and Embedding} (a) A ring of antiferromagnetically arranged spins with a domain wall (red) comprising aligned neighboring pair of spins. (b) Real space rendering of a domain wall distributions at  high($\Jen=1$) and low ($\Jen=0.1$) coupling in two configurations measured from the QPU for the same system size.
    (c) The empirical domain wall probability distribution (bar plots) obtained from annealer measurements along with the most similar theoretical domain wall probability distribution (line plots) is shown for two coupling strength: $\Nqb = 301$, $\tau = 10 \mu s$, on \texttt{Advantage2\_System1.1}, $\Jen=0.75,0.25$). Note the poor agreement at strong couplings, small times, for this machine, discussed in the text. 
    (d) Empirically domain-wall density $\langle \ndw \rangle$ (dots) at different extracted effective temperatures $\Te$ for \texttt{Advantage\_system4.1} as shown on top of the analytical expressions (solid lines) over a range of system sizes. (e) An embedding of a 1D ring onto D-Wave's Zephyr Architecture, without chaining, one qubit per spin.}
    \label{fig:fig1}
\end{figure*} 

Here we test the assumption that the system produces a classical Gibbs ensemble on QAs at large qubit scale, by benchmarking experimental results with an exact model: the transverse quantum Ising ring~\cite{mbeng2024quantum}. 
We perform experiments on four different D-Wave quantum processing units (QPUs), with system size $N_{\text{qb}}$ spanning three orders of magnitude up to 4000 qubits, and across a range of coupling constants $J$ and annealing times $\tau$. 
We extract an effective, dimensionless sampling temperature $t_{\text{eff}}$ from the data and plot it against the annealing parameters.
Finally we relate this effective temperature to a physical temperature and compare it with the machine temperature. 

\section*{Experiment}

D-Wave's QAs comprise superconducting flux qubits connected to each other via programmable couplers in hardware-specific graphs.
We program a one-dimensional Hamiltonian of the form
\begin{align}
        {\mathcal H} =& - \frac{A(s)}{2} \sum_i \hat\sigma^x_{i} 
     + \frac{B(s)} {2} \Jen
        \sum_{i<j}  \hat\sigma^z_{i} \hat\sigma^z_{i+1}
     \label{equation:QA_Hamiltonian_h_gain}
\end{align}
where $\Jen\in [0,1]$ is the ``encoded coupling", a dimensionless quantity programmed into the machine (all dimensionless quantities will be in lower case), whereas the annealing functions $A(s)$ and $B(s)$ are hardware-dependent \textit{energy scales} that vary monotonically with the annealing parameter $s \in [0,1]$, with $A(s)$ decreasing to zero and $B(s)$ increasing from approximately zero to a maximum value with increasing $s$. 
The corresponding \emph{physical} coupling at the end of the anneal is therefore
\begin{equation}
\Jp = {B(1) \Jen}/{2}.
\label{eq:Jp}
\end{equation}
The values of $B(1)$ are of the order of magnitude of a fraction of a Kelvin, for example, $B(1)\approx 0.407$ K for the \texttt{Advantage\_system4.1}~\cite{dwave-docs}. (Note that in this work {\em We measure all the energies in unit of temperature.}) Values for other QPUs are of similar order or magnitude, with more recent chips having larger values for more flexibility. Finally, note also that it has been pointed out that qubits might ``freeze" before the end of the anneal~\cite{AminSearching2015}, leading to a potentially lower $\Jp$. Here we assume Eq.~\eqref{eq:Jp} to hold. 

The QPUs are maintained at cryogenic temperatures $\Tm \approx 15$ mK~\cite{dwave-docs} (which can vary slightly from one qubit to another).
If that is the case, then one expects the ensemble to be governed by the ratio $\Tm/ \Jp$.
Because $\Tm$ is fixed by the hardware, the thermal ensemble can be varied by changing the  \textit{energy scale} parameter $\Jen$~\cite{RaymondGlobal2016,KingQubit2021,lopez2024quantum,SatheClassical2025,TezaFinitetemperature2025}.
A \emph{dimensionless effective temperature} (i.e. a temperature measured in the energy scale of the model being studied) was often assumed~\cite{KingQubit2021,Lopez-BezanillaKagome2023} in the form
\begin{equation}
    \Te \propto \Jen^{-1}.
    \label{eq:Teff0}
\end{equation}
A better heuristic form which takes into account the influence of annealing time $\tau$ and system size $\Nqb$ might be guessed to be
\begin{align}
\Te = \alpha\left(\tau, \Nqb \right) \frac{\Tm}{\Jp},
\label{Eq:Teff}
\end{align}
where $\alpha\left(\tau, \Nqb\right)$ is a machine-dependent dimensionless pre-factor. 
Intuitively, one expect that $\alpha$ would depend only weakly, if at all, on size $\Nqb$, while depending strongly on annealing time, monotonically decreasing with $\tau$ to a limit value because longer anneals yield lower-energy configurations. In the opposite regime, when timescales become comparable to the system’s coherence time (tens to hundreds of nanoseconds in contemporary hardware), the output follows quantum Kibble–Zurek scaling~\cite{KingCoherent2022}---see, however, Ref.~\cite{GyhmBoltzmann2025}, which suggests a classical Boltzmann sampling even with closed quantum evolution in the diabatic limit.
The transition between these two extremes remains yet unexplored. 

Besides the machine temperature $\Tm$ and the effective temperature $\Te$, a third temperature is relevant: the {\em physical} temperature $\Tp$ defined as the temperature that the qubits should have in order to return the observed ensemble given the physical coupling, in a thermalized classical system, or 
\begin{equation}
 \Tp=\Jp \Te.
 \label{Eq:Tp}
\end{equation}
Then, from Eqs.~(\ref{Eq:Teff},\ref{Eq:Tp}), it follows that
\begin{equation}
\alpha\left(\tau, \Nqb \right)=\frac{\Tp}{\Tm}.
\label{eq:alpha_ratio}
\end{equation}
As the ratio between physical to machine temperatures, one would expect $\alpha$ to converge to unity for sufficiently long annealing times.

We test these notions on the one-dimensional periodic antiferromagnetic Ising ring of odd length $\Nqb$, described by the dimensionless Hamiltonian
\begin{align}
h = \sigma_1^z \sigma_2^{z}+ \sigma_2^z \sigma_3^{z}+\dots+\sigma_{\Nqb}^z \sigma_{1}^{z},
\label{eq:h}
\end{align}
which admits an exact solution and can be natively embedded (i.e., with one logical spin implemented by one physics qubit in the QPU) on D-Wave’s Advantage processors (see Fig.~\ref{fig:fig1}e). 

A domain wall is a couple of neighboring spins pointing in the same direction, thus breaking the antiferromagnetic order, as shown schematically in Fig.~\ref{fig:fig1}a. 
The probability distribution for the number of domain walls $\Ndw$ is known to be:
\begin{align}
p(\Ndw;\Te) = 2 \binom{\Nqb}{\Ndw} \frac{e^{- 2\Ndw/\Te}}{Z(\Te, \Nqb)}.
\label{eq:domain_wall_probability}
\end{align}
where  $Z(\Te, \Nqb)$ is the partition function, reported as Eq.~\eqref{eq:partition} in the Methods section. $\Nqb$ is always odd, and so is $\Ndw$, assuring at least one domain wall. 
 
\begin{figure*}[t!]
    \includegraphics[width=0.99\textwidth]{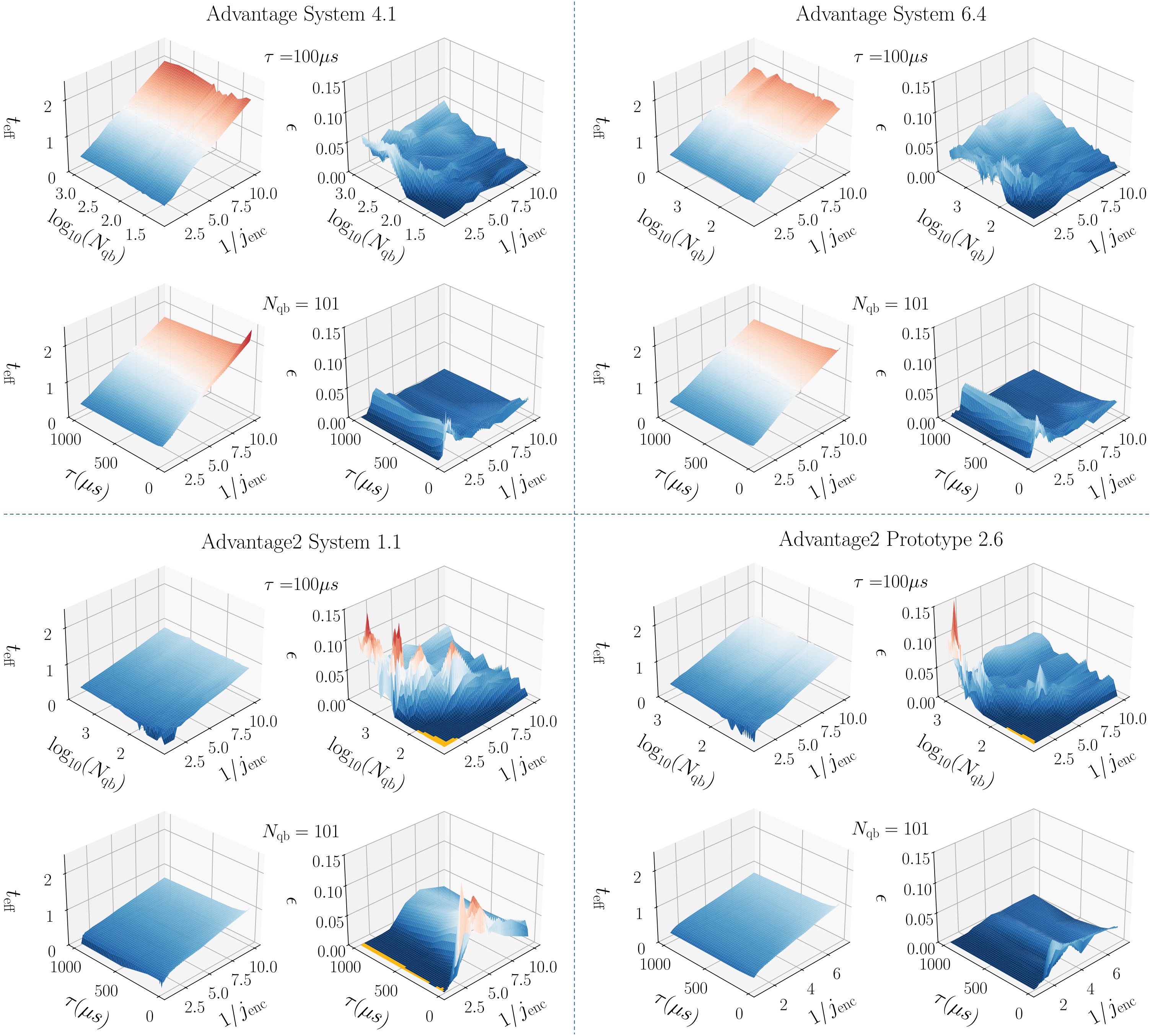}
   \caption{{\bf Effective Temperatures and TVD Deviations from Gibbs Distributions} for the four machines. Each sub-figure corresponds to a machine and contains 3D Surface plots of the effective temperature ($\Te$), and TVD ($\epsilon$) ploted versus the inverse encoded energy coupling $1/\Jen$ and order of magnitude of system size $\log_{10}(\Nqb)$ or annealing time $\tau$. The small gold colored regions in the plots for $\epsilon$ for Advantage2 devices correspond to zero temperature samples, for which $\epsilon$ is not plotted: as explained in the text, strong energy couplings and small system sizes lead to inapplicable sampling distributions and a breakdown of our thermometry methods, because the ground state is always reached.}
    \label{fig:surface_plot}
\end{figure*}

\begin{figure*}[t!]
     \centering
     \begin{subfigure}[b]{0.45\textwidth}
         \centering
         \includegraphics[width=.9\textwidth]{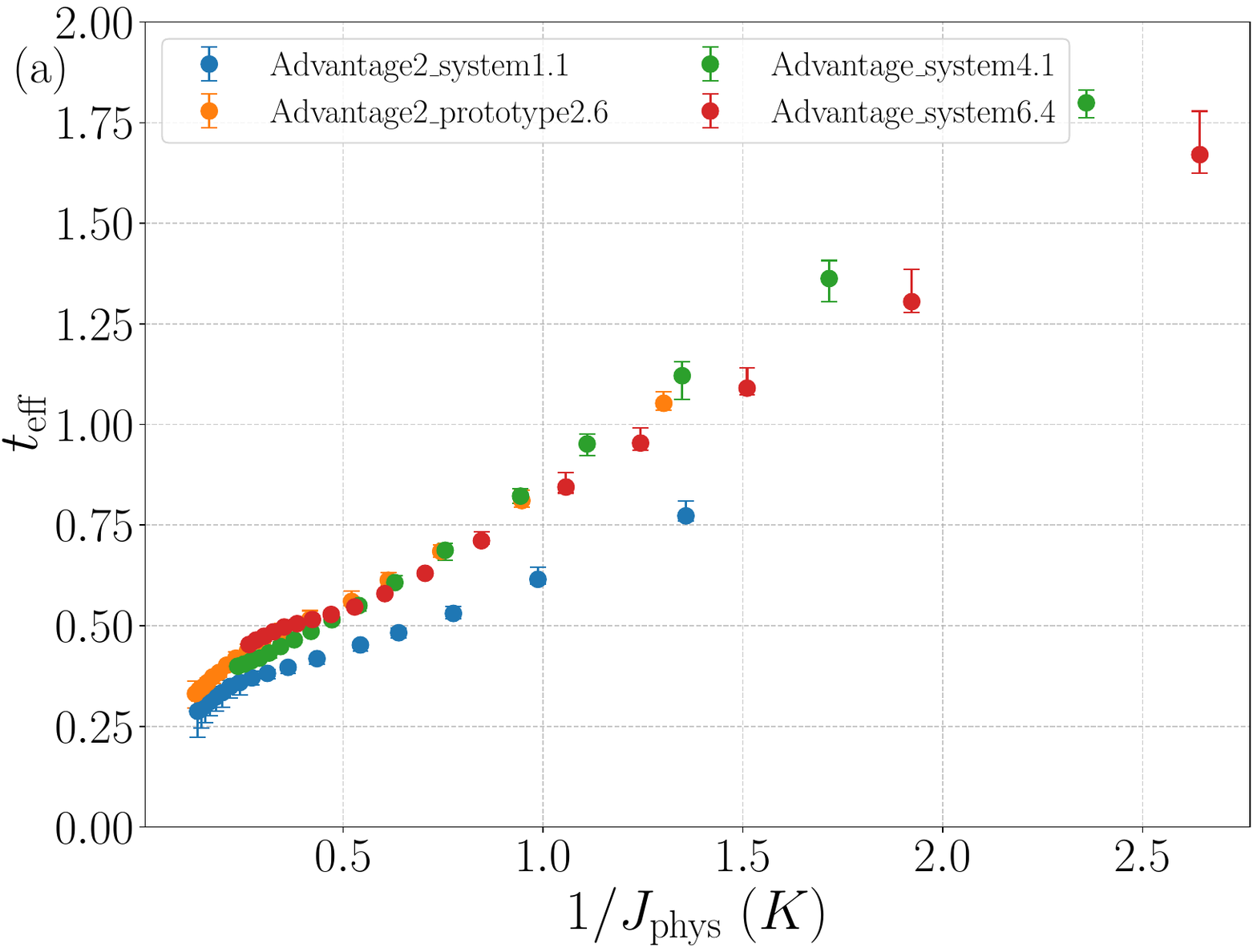}
     \end{subfigure}  
     \begin{subfigure}[b]{0.45\textwidth}
         \centering
         \includegraphics[width=.9\textwidth]{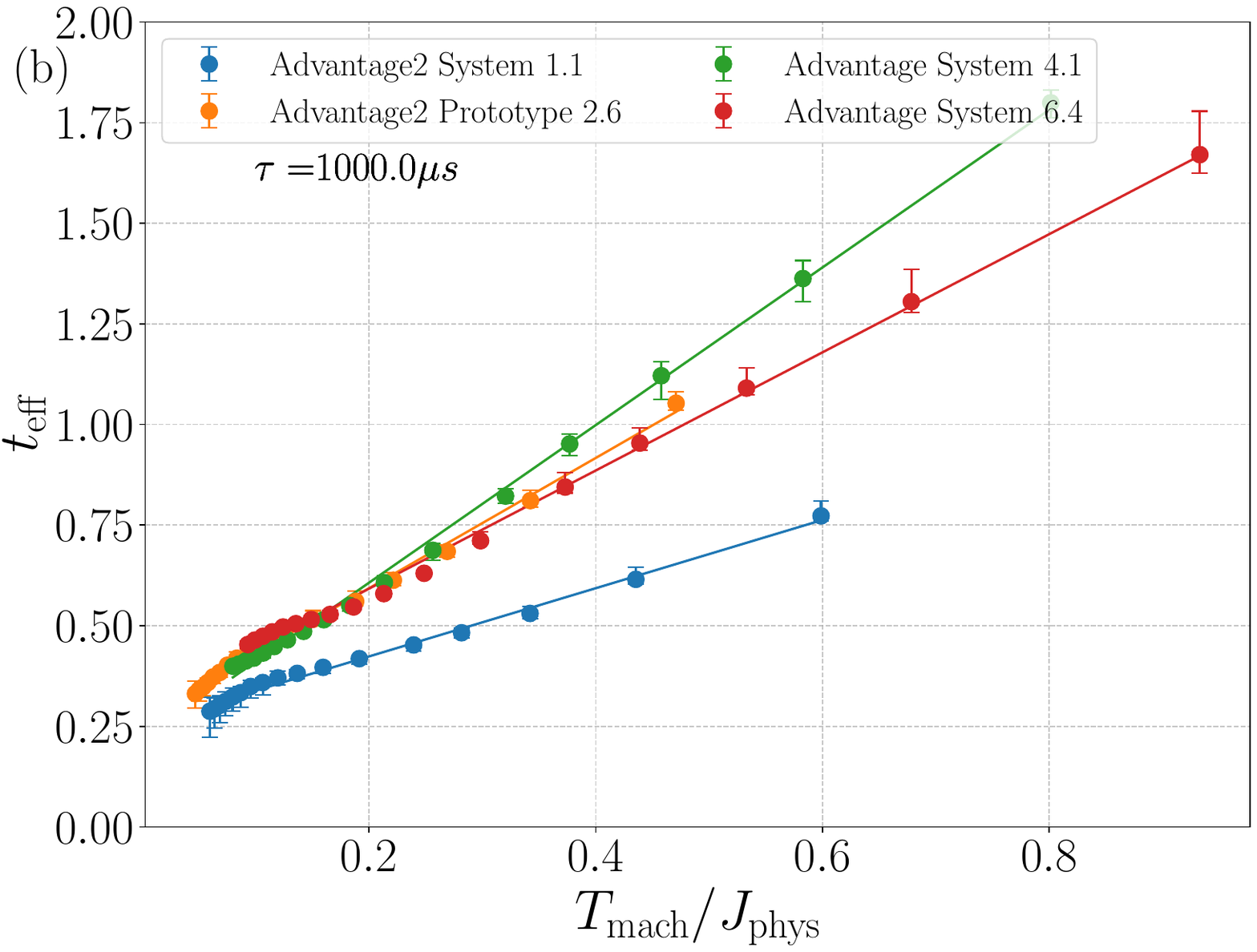}
         
     \end{subfigure}
     \begin{subfigure}[b]{0.45\textwidth}
         \centering
         \includegraphics[width=.9\textwidth]{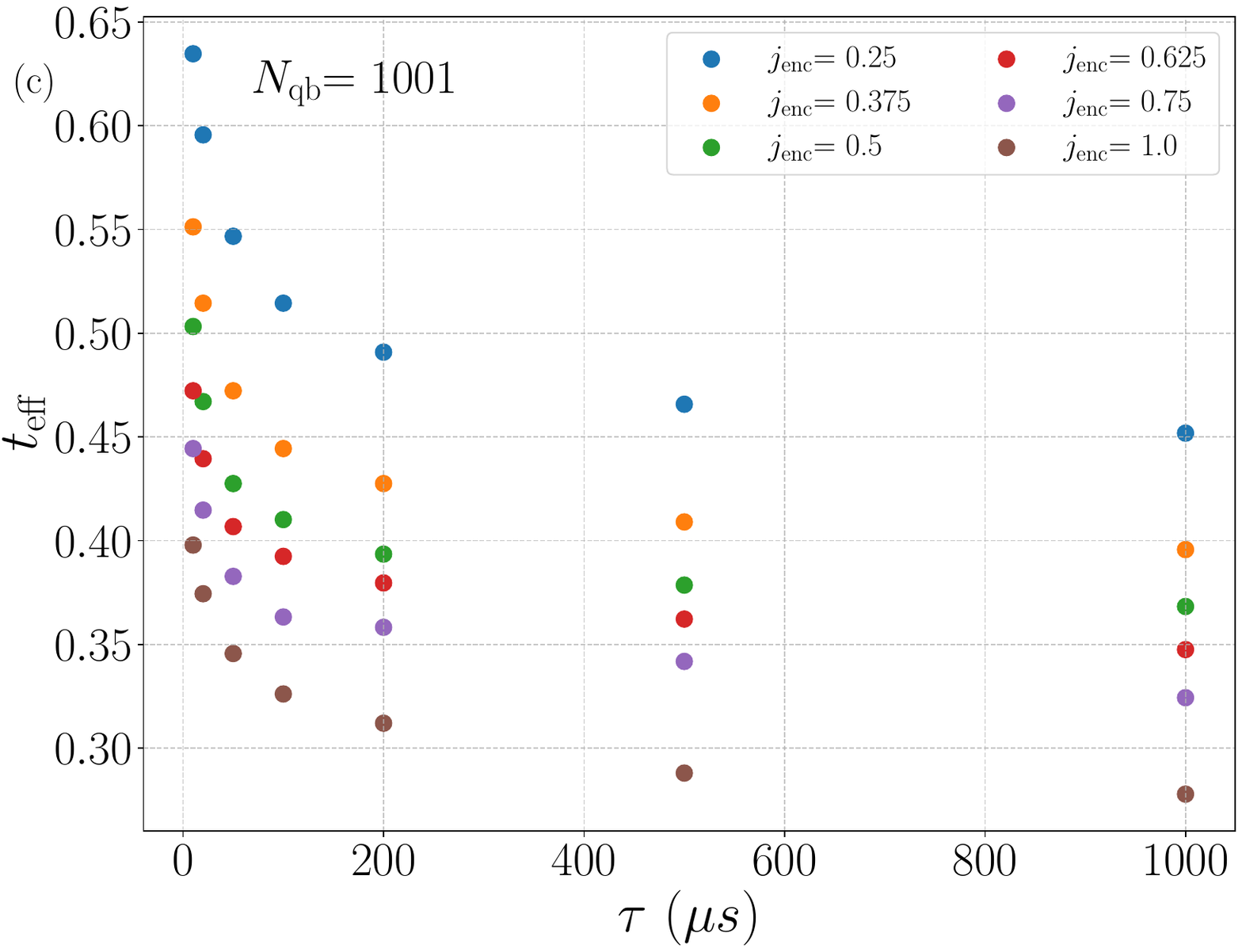}
     \end{subfigure}
     \begin{subfigure}[b]{0.45\textwidth}
         \centering
         \includegraphics[width=.9\textwidth]{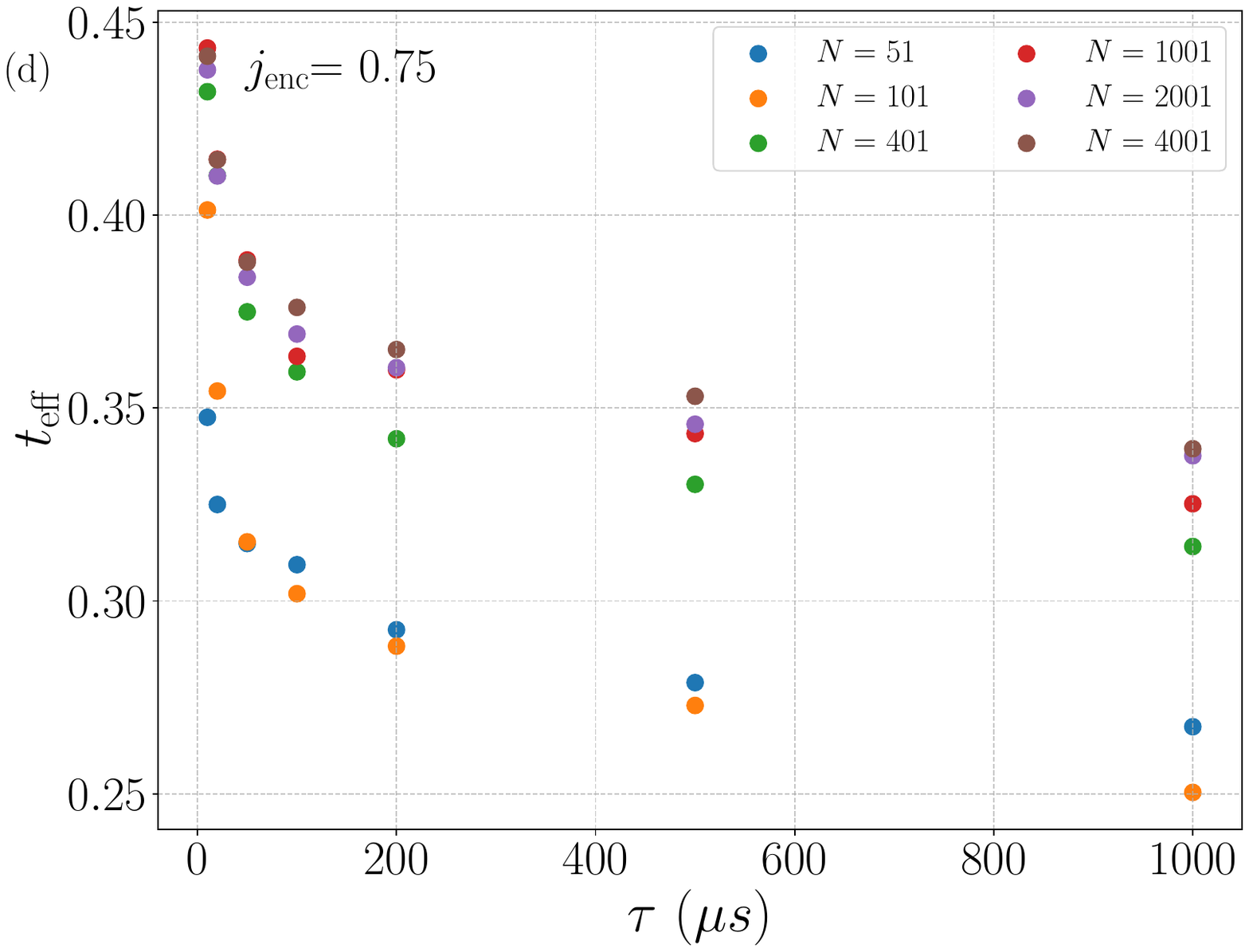}
     \end{subfigure}
    \caption{{\bf Effective Temperature $\Te$} plotted vs. $1/\Jen$ (a) and $\Tm/\Jp$ (b) for the four machines. The data points are averaged over sizes $\Nqb \ge 100$, and the error bar shows the very small fluctuation in $\Te$ across different sizes $\Nqb$, with the lower (upper) bound corresponding to the smaller (larger) $\Te$. Note that data across different machines collapse around similar at the same physical coupling $\Jp$, with the exception of \texttt{Advantage2\_system1.1}. The slope of panel (b) is the dimensionless ($\Jp$ is measured in units of temperature) constant $\alpha$ of Eq.~(\ref{Eq:Teff},\ref{eq:alpha_ratio},\ref{Eq:Teff2}). Plots of $\Te$ vs. $\tau$ on QPU \texttt{Advantage2\_system1.1} programmed at $\Jen=0.75$ for difference values of $\Nqb$ (c) or for $\Nqb=1001$ and different values of $\Jp$ (d) show monotonic decrease of the effective temperature converging to asymptotic values at large annealing time, .}
    \label{fig:cuts}
\end{figure*}

We perform forward annealing runs across a range of system sizes $\Nqb$, coupling constants $\Jen$, and annealing times $\tau$ on four machines. The three parameters $\Nqb, \Jen, \tau$  define distinct ensembles, from each of which we acquire ${\sim}10,000$ measurements.
From the measured domain-wall configurations (Fig.~\ref{fig:fig1}b) we construct the empirical distribution $\xi(\Ndw)$ (Fig.~\ref{fig:fig1}c). 
We extract the effective temperature $\Te$ by minimizing the total-variation distance (TVD)
\begin{align}
\epsilon(\Te) = \frac{1}{2}\sum_{\Ndw} \abs{\xi(\Ndw) - p(\Ndw;\Te)},
\label{eq:tvd}
\end{align}
between the observed distribution $\xi(\Ndw)$ and the theoretical prediction $p(\Ndw)$ given by Eq.~\eqref{eq:domain_wall_probability}, using a conjugate gradient algorithm.
The TVD deviation $\epsilon$ ranges from $0$ to $1$ and  is of intuitive interpretation as a ``maximal percentage error". 
The initial estimate in our conjugate gradient approach is obtained by extracting an effective temperature from the recorded mean domain-wall density $\langle \ndw \rangle =  \langle \Ndw \rangle / \Nqb$, whose dependence on $\Te$ is obtained exactly from Eq.~\eqref{eq:domain_wall_probability} and reported in Eq.~\eqref{eq:domain_wall_average} in the Methods section.
Figure~\ref{fig:fig1}c shows two representative fits for \texttt{Advantage2\_System1.1} samples obtained for $\Nqb = 301$, $\tau= 10 \ \mu$s, and two values of $\Jen$: at $\Jen=0.75$, we observe stronger deviations at small $\tau$ and large $\Jen$ (a situation we discuss more later); and at $\Jen=0.25$ which shows a better agreement. 

Any spin-based methodology breaks down at high $\Te$, because the disorder is bounded by the maximum entropy per spin, $\ln 2$. Beyond a certain temperature, observables saturate, and  statistical sampling becomes ineffective. 
A second limitation arises at low $\Te$ (i.e., low coupling $\Jp$) and small size $\Nqb$ because the ground state is reached at finite temperature. Consider Fig.~\ref{fig:fig1}d, which shows the theoretical domain wall density $\langle n_{\text{dw}} \rangle$ as a function of $\Te$, alongside experimental data from forward anneals on \texttt{Advantage\_system4.1} at $\tau = 10\,\mu$s. The data collapses across different sizes, except at very low $\Te$ where the domain-wall density saturates to its ground state value of $1/\Nqb$ (a single domain wall). For instance, looking at the figure, for $\Nqb<101$ one expect the method to fail for $\Te<0.4$. 

\section*{RESULTS}
Our results confirm overall the Gibbsianization assumptions via our heuristic formula, but with some important caveats.  Figure~\ref{fig:surface_plot} shows surface plots of the effective temperature $\Te$ and TVD error $\epsilon$ for the four machines, as functions of reciprocal coupling $1/\Jen$, system size $\Nqb$, and annealing time $\tau$. Note that effective temperature appears to be largely independent of $\Nqb$, an assumption critical to finite-size scaling analyses in prior studies~\cite{SatheClassical2025,TezaFinitetemperature2025}. 
This contrasts with some prior temperature estimation studies which considered smaller systems and found a significant system size dependence~\cite{RaymondGlobal2016}.
Fluctuations of $\Te$ in $\Nqb$ are visible at high $\Te$ (typically not an interesting thermal region) particularly on older QPUs (top row), where they may also result from limitations of our methodology, as discussed above. These fluctuations are much higher at shorter annealing times, and significantly reduced in the surfaces corresponding to longer annealing times: see Supplementary Figures 1-7 for $\tau$ ranging from 10 to 1000 $\mu$s.

The surface plots of $\Te$ are, with good approximations, planar, consistent with Eq.~\eqref{eq:Teff0}--but see below for important correction.
This is shown more clearly in Fig.~\ref{fig:cuts}a where we plot $\Te$ vs. $1/\Jen$ averaged over all system sizes with $\Nqb >200$, and we observe linear dependence, with stronger deviations at low temperature where our methodology becomes less appropriate (see above).

Significantly, later generations machines---\texttt{Advantage2\_system1.1} and \texttt{Advantage2\_Prototype2.6}, (bottom row of Figure~\ref{fig:surface_plot})---return considerably lower  effective temperatures across all annealing parameter ranges, by a factor ${\sim}2$---See Fig.~\ref{fig:surface_plot} and Fig.~\ref{fig:cuts}a,b---consistent with higher energy scales in these machines for greater success probability in optimization problems. These machines reach an effective temperature below the crossover discussed above, where our methodology fails because the annealer easily finds the ground state. That is shown by the drop of $\Te$ at small $\Nqb$ and $1/\Jen$ in Fig. \ref{fig:surface_plot}, and is marked by a gold region in the surface plots of $\epsilon$.

Surface plots of the TVD---$\epsilon$ in Eq.~(\ref{eq:tvd})---qualify the fidelity of Gibbsianization   and are reported as second column figures within each of the four sub-panel of Figure~\ref{fig:surface_plot}). Older machines (top row) consistently show  low values of $\epsilon$ ($\lesssim 5\%$), with few exception at longer annealing times (see also supplementary Figures 1-3) indicating  a high fidelity of Boltzmann sampling across all devices. There, the error tends to increase with size, although its features are complicated. 

Importantly, the newer generation, higher power  \texttt{Advantage2} machines show a larger error, suggesting that the qubits are not thermalizing with the environment. This could be due to greater quantum coherence~\cite{matsuda2009quantum} manifesting at small annealing times. Indeed, \texttt{Advantage2\_System1.1} exhibits significantly larger deviations from Gibbs sampling at short annealing times ($\simeq\mu$s) where coherence is larger (see supplementary Figures 4-6). 

When $\Te$ is plotted against the inverse physical coupling $1/\Jp$ in Fig.~\ref{fig:cuts}b (with $\Jp$ in units of $\Tm$, data taken at $\tau = 1000\,\mu$s), effective temperatures nearly collapse across three machines at the same $\Jp$. The newer \texttt{Advantage2\_System1.1} is again exceptional, exhibiting systematically lower temperatures at equivalent coupling. 

Overall, we have observed that the  intuitive predictions are corroborated by the data: 1) the ensembles follow a Gibbs distribution with excellent approximation across orders of magnitude in size; 2) $\Te$ scales linearly with the reciprocal coupling constant; 3) $\Te$  shows little dependence on $\Nqb$, with the noted exceptions; 4) $\Te$ varies with annealing time in expected ways monotonically decreasing to an asymptotic value as $\tau$ increases, as shown in Figs.~\ref{fig:cuts}c,d.

\begin{figure*}[t!]
    \centering
    \begin{subfigure}[b]{0.49\textwidth}
    \centering
    \includegraphics[width=0.99\textwidth]{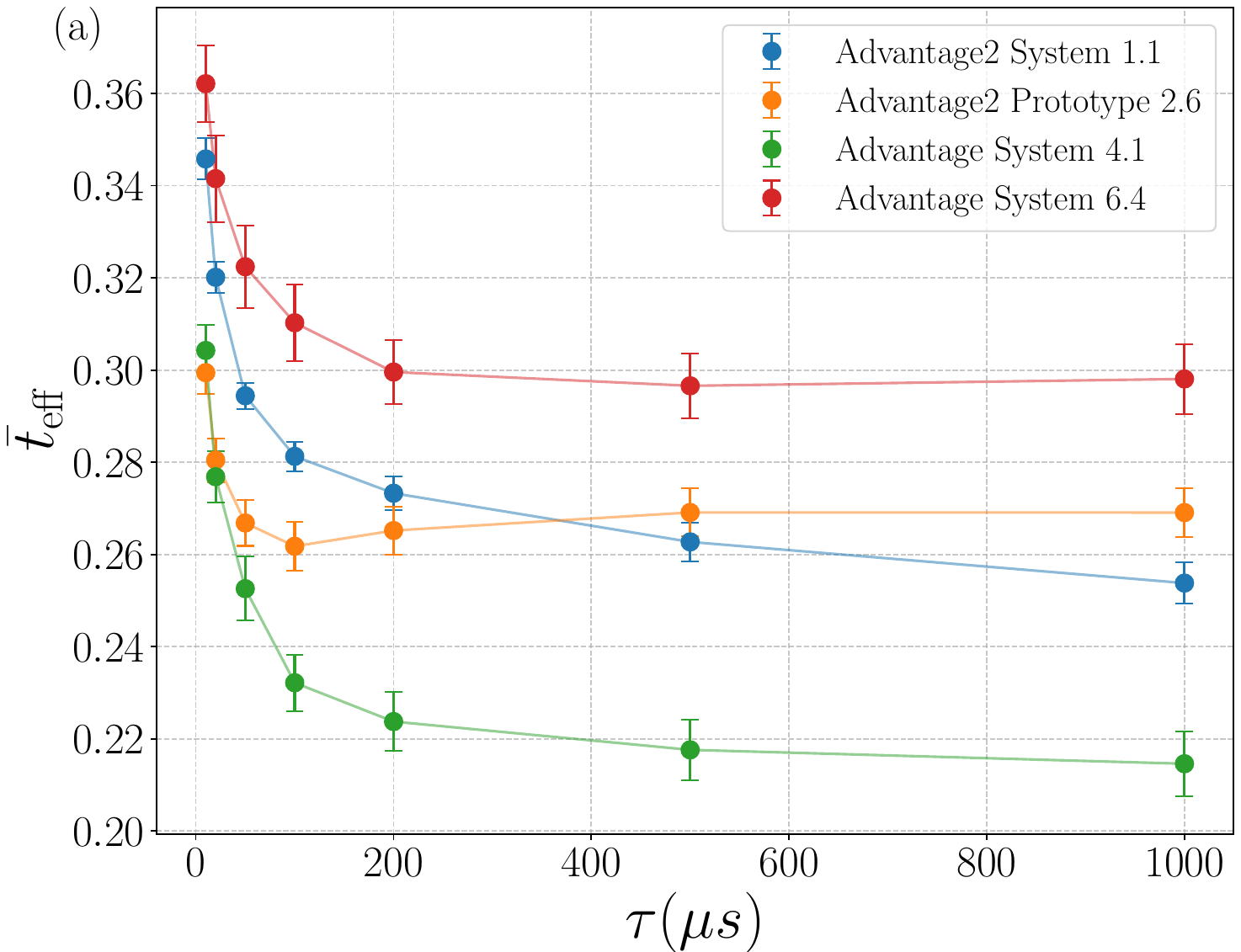}
    \caption{}
    \end{subfigure}
    \hfill
    \begin{subfigure}[b]{0.49\textwidth}
    \centering
    \includegraphics[width=0.99\textwidth]{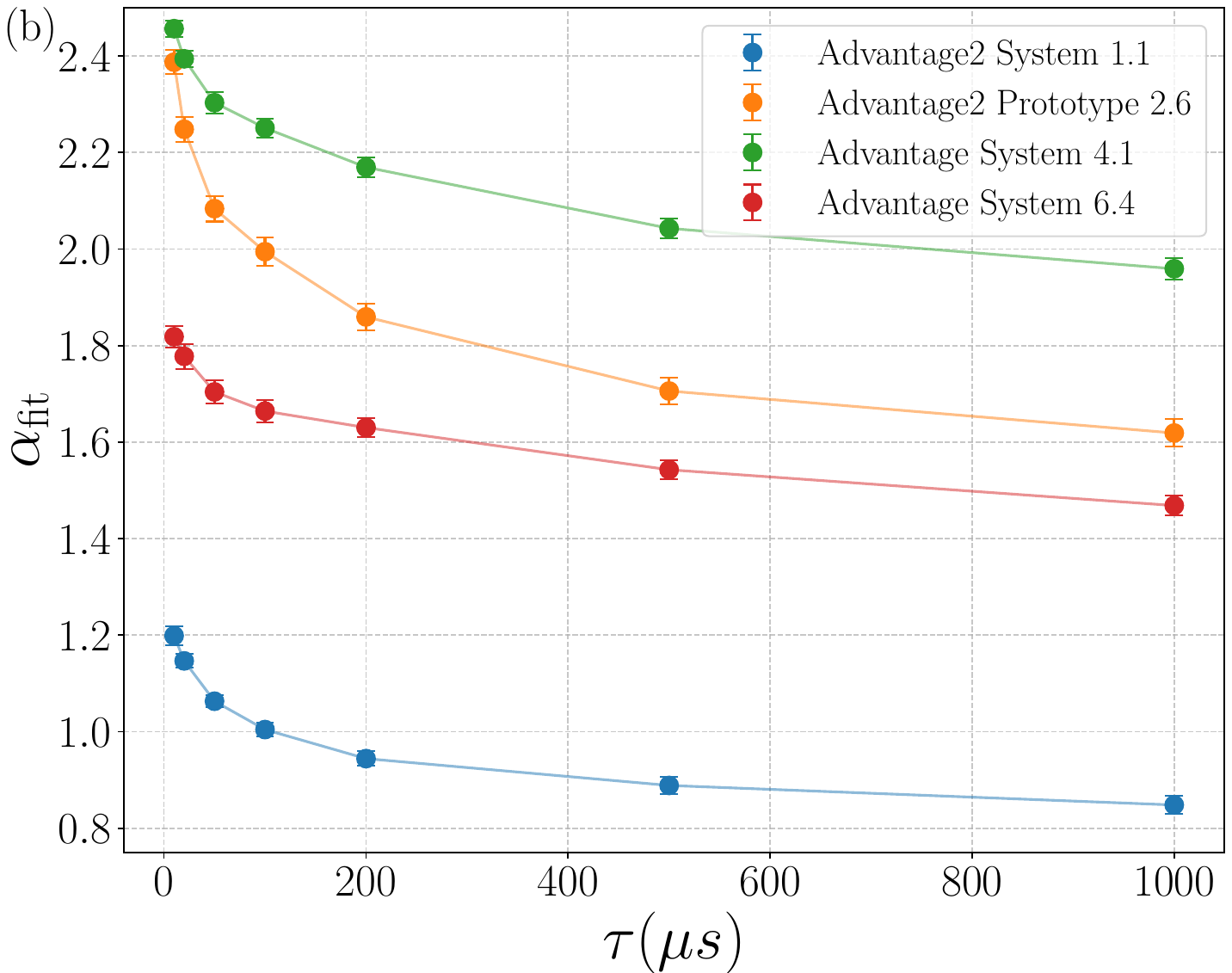}
    \caption{}
    \end{subfigure}
    \caption{{\bf Extracted Parameters for the Effective Temperature} The temperature offset $\Teb$ (a) and the slope $\alpha$ (b) obtained from fitting Eq.~(\ref{Eq:Teff2}) to data from annealing $T$ for all four machines, are plotted vs. the annealing time $\tau$. As in Fig..~\ref{fig:cuts}b,c, the values are monotonically decreasing in annealing time, saturating at long annealing.}
    \label{fig:fig4}
\end{figure*}

There is, however, one very significant deviation from the initial assumptions. Returning to Fig.~\ref{fig:cuts}b, the slope of the linear fit allows for the extraction of $\alpha$ according to Eq.~(\ref{Eq:Teff}). Remarkably, the fits do not support Eq.~(\ref{Eq:Teff}), previously assumed in several works~\cite{lopez2024quantum,KingQubit2021,Lopez-BezanillaKagome2023,KingObservation2018,AminSearching2015,SatheClassical2025} and instead exhibit a nonzero offset $\Teb$:
\begin{align}
\Te = \Teb + \alpha\left(\tau, \Nqb \right) \frac{\Tm}{\Jp}.
\label{Eq:Teff2}
\end{align}
The temperature offset is non-negligible, ranging from $0.2$ to $0.4$ across machines and annealing times. This temperature floor $\Teb$ is independent of $\Jp$ but depends on annealing time, as shown in Fig.~\ref{fig:fig4}a, while Fig.~\ref{fig:fig4}b displays the corresponding values of $\alpha$ extracted from fits. In other words, the ensembles are Gibbs-distributed, but in future applications involving sampling, the commonly assumed Eq.~\eqref{Eq:Teff} could be replaced by Eq.~\eqref{Eq:Teff2}.

The presence of this offset complicates the notion of a physical temperature, as introduced in the discussion leading to Eq.~(\ref{Eq:Tp}), where $\Tp$ is defined as the temperature the qubits would need in order to produce the observed Gibbs distribution at physical coupling $\Jp$. Substituting Eq.~(\ref{Eq:Teff2}) into Eq.~(\ref{Eq:Tp}) yields a coupling-dependent physical temperature, which would undermine the logic of the original argument.
Alternatively, if $\Teb$ is viewed as a constant shift in the dimensionless effective temperature,  it might be a signature of residual uncorrelated longitudinal or flux noise affecting individual qubits and which is unaffected by the coupling.
If this interpretation holds true, then once could perhaps replace Eq.~\eqref{Eq:Tp} with $\Tp = \Jp \left(\Te - \Teb\right)$, in which case $\alpha$ would retain its meaning as the ratio between the physical and machine temperatures, as in Eq.~\eqref{eq:alpha_ratio}. As shown in Fig.~\ref{fig:fig4}b, this ratio decreases with annealing time and remains above unity, as expected, except for one case: \texttt{Advantage2\_System1.1}, which again deviates from the general trend, with $\alpha$ dropping below one. This may reflect the distinct character of this processor, which, as noted above multiple times, and illustrated in Fig.~\ref{fig:cuts}a,b, features higher power and stronger coherence than other devices, leading to lower effective temperatures and higher deviation from Gibbsianization. Alternatively, one may even question the utility of our (re)definition of $\Tp$. We consider it operationally meaningful insofar as it yields consistent and interpretable results on machines that perform as reliable classial  Gibbs samplers. Moreover, the value of $\Tp$ depends on the value of $\Jp$ and we have already mentioned how Eq.~\eqref{eq:Jp} would need to be modified if one supposes a so-called ``freeze-out" mechanism~\cite{AminSearching2015}. We leave all these questions open for future works, which will have to investigate the origin of $\Teb$ within the context of the quantum annealing process. 

Finally, if $\Teb$ originates from noise, it may be both model-dependent and correctable through improved calibration~\cite{ChernTutorial2023}. No such calibration was applied in this study, in order to maintain generality and avoid dependence on machine-specific tuning protocols. However, future studies should examine whether $\Teb$ varies with calibration procedure or the choice of embedded Hamiltonian. 

\section*{CONCLUSION}
Understanding annealed ensembles, their thermalization, distribution, coherence, or lack thereof, is essential for many potential uses of quantum annealers, whether as analogue classical or quantum simulators, or as non-local optimizers, and samplers in machine learning applications. We have shown that key assumptions about Gibbsianization and effective-temperature control generally hold in large-scale systems where they had never been tested, but only within specific parameter regimes and with notable qualifications.

Importantly, we confirm Gibbsianization with an effective temperature $\Te$ that decreases with annealing time $\tau$ and scales inversely with encoded coupling $\Jen$, aligning prior heuristic, intuitive assumptions. Moreover, we verify that for most machines, the unitless effective temperature is largely independent of system size $\Nqb$ (in contrast to some prior work on temperature estimation), with deviations becoming more pronounced at smaller couplings and smaller annealing times (i.e. larger effective temperatures). This supports the use of these devices for finite-size scaling analyses that have been successfully used for studying classical statistical physics models. However, our results also reveal significant deviations: most notably, the widely assumed relation $\Te \sim 1/\Jen$ fails due to a non-negligible coupling-independent offset likely due to machine noise. We also find that newer-generation QPUs display larger deviations from Gibbs distributions, likely reflecting increased quantum coherence and reduced thermalization efficiency.

Our findings represent the first example of scalable, physically grounded approach to thermometric benchmarking of quantum annealers, across three orders of magnitude in system size. Future work should explore the transportability of this thermometry across models (e.g., using the 1D Ising ring as a ``thermometer'' to predict effective temperatures in more complex systems), how fine calibration affects the temperature offset, whether the reported Gibbsianization holds in large scale frustrated systems, especially on more coherent machines where transverse fields could split classical ground states, and if any role is played by a freeze out mechanism. 

\section*{METHODS}

We ran our experiments on four D-Wave quantum annealers:
\texttt{Advantage\_system4.1}, \texttt{Advantage\_system6.4}, \texttt{Advantage2\_System1.1},  \texttt{Advantage2\_Prototype2.6}.

All experiments used a standard forward annealing schedule (no reverse annealing), with automatic rescaling disabled, flux drift compensation enabled, and $10\,\mu$s of readout thermalization. We explored a wide range of annealing times $\tau$, encoded coupling strengths $j_{\text{enc}}$, and system sizes $\Nqb$, with odd values ranging from $11$ to $4001$. Each system size was embedded without chaining (i.e., one physical qubit per logical spin).

For each combination of parameters $(\tau, \Jen, \Nqb$, we collected between 10,000 and 100,000 annealing runs (shots), depending on the difficulty of the regime. In high-coupling or low-temperature regimes—where statistical resolution is poorer—we collected up to 100,000 samples per point to ensure reliable temperature extraction. More moderate regimes generally required only 10,000–20,000 samples.

To manage QPU resource limits (1 second per submission), we dynamically allocated sample counts using linear interpolation based on the expected experimental complexity. For instance, experiments with longer annealing times or stronger couplings were assigned more samples. We estimated execution time from anneal duration, programming time, readout, and thermalization overheads, and constructed sampling schedules that maximized coverage within the QPU access window.

Once sampling data were obtained, we extracted domain-wall counts $\Ndw$ from the raw data and constructed a normalized histogram $\xi(\Ndw)$ for each parameter combination. We then extracted the effective temperature $\Te$ by minimizing the total variation distance TVD [Eq.~ \eqref{eq:tvd}] between the observed distribution and the exact probability predicted distribution in Eq.~\eqref{eq:domain_wall_probability}, whose partition function is
\begin{align}
    \begin{split}
       Z(\Te, \Nqb) &=  2^{\Nqb} \exp \left(-\Nqb/\Te\right) \times \\ 
    &[ \cosh^{\Nqb}\left(\Te^{-1}\right) - \sinh^{\Nqb}\left(\Te^{-1}\right) ],
    \end{split} 
 \label{eq:partition}
\end{align}

Minimization was carried out using a conjugate gradient algorithm with line search. As an initial guess, we inverted the exact expression for the average domain-wall density $\langle \ndw \rangle = \langle \Ndw \rangle / \Nqb$, derived from the partition function:

\begin{align}
\begin{split}
\langle \ndw \rangle  =  \frac{1}{2} & \biggl\{  1 - \frac{\sinh (\Te^{-1}) \cosh(\Te^{-1})}{\cosh^L(\Te^{-1}) - \sinh^L (\Te^{-1})} \times \\
&\left[\cosh^{\Nqb-2} (\Te^{-1}) - \sinh^{\Nqb-2} (\Te^{-1})\right] \biggl\}.
\end{split}
\label{eq:domain_wall_average}
\end{align}

This optimization converged reliably in under 30 iterations per parameter set. All data processing and annealing submissions were handled through the Ocean SDK API via D-Wave’s Leap platform.

Heuristic techniques to improve the sampling quality (such as gauge averaging and calibration refinement of the coupler strengths and flux bias offsets) were not applied in this study to retain architecture-agnostic comparability.

\section*{Acknowledgments} 

We thank Andrew King for helpful discussions.
We acknowledge the support of NNSA for the U.S. DOE at LANL under Contract No. DE-AC52-06NA25396, and Laboratory Directed Research and Development (LDRD) for support through 20240032DR.
PS also acknowledges the support from the National Security Education Center (NSEC) Informational Science and Technology Institute (ISTI) using the Laboratory Directed Research and Development program of Los Alamos National Laboratory project number 20240479CR-IST.
LANL is managed by Triad National Security, LLC, for the National Nuclear Security Administration of the U.S. DOE under contract 89233218CNA000001. \\

\section*{ Author Information}

These authors contributed equally: George Grattan and Pratik Sathe. \\

\textbf{Contributions.} 
The three authors conceived of the project together. 
% GG and PS implemented it on D-wave machines.
GG ran experiments on the machine, and performed data analysis.
CN and PS provided the theoretical approach. All authors provided data interpretation and suggested experiments. CN supervised the project and wrote the draft. All authors contributed to the final manuscript. \\ 

\textbf{Corresponding authors.}
Correspondence should be sent to all the authors \\

\section*{Ethics Declaration}
\textbf{Competing interests.} 
PS is an employee of D-Wave Quantum Inc. and declares competing interests on that basis.
The other authors declare no competing interests.

\setcounter{equation}{0}
\setcounter{figure}{0}
\setcounter{section}{0}
\setcounter{table}{0}
\setcounter{page}{1}
\makeatletter
\renewcommand{\theequation}{S\arabic{equation}}
\renewcommand{\thepage}{S-\arabic{page}}
\renewcommand{\thesection}{S\arabic{section}}
\renewcommand{\thefigure}{S\arabic{figure}}

\begin{widetext}
\section*{Supplemental Material to ``Classical Thermometry of Quantum Annealers''}

\begin{figure*}[h!]

    \centering
    \begin{subfigure}[b]{0.48\textwidth}
        \centering
        \includegraphics[width=\textwidth]{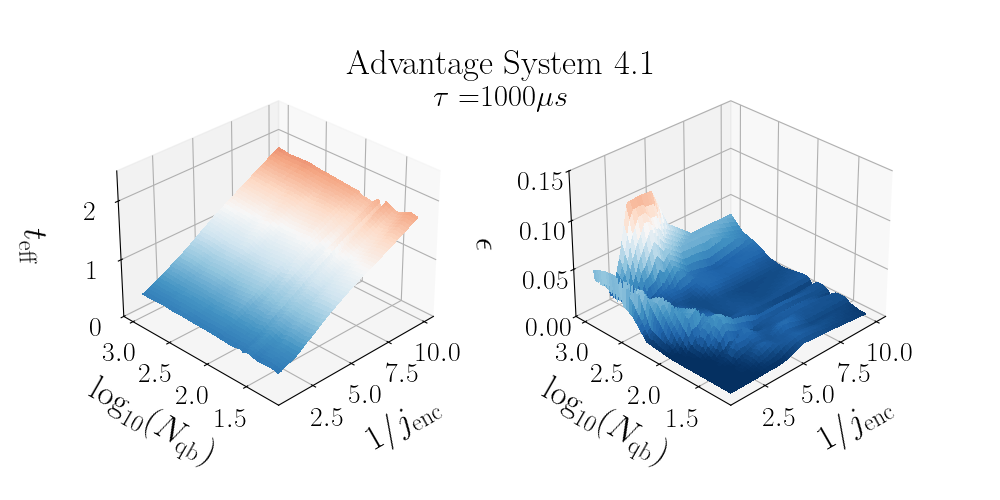}
        \caption{}
    \end{subfigure}
    \quad
    \begin{subfigure}[b]{0.48\textwidth}
        \centering
        \includegraphics[width=\textwidth]{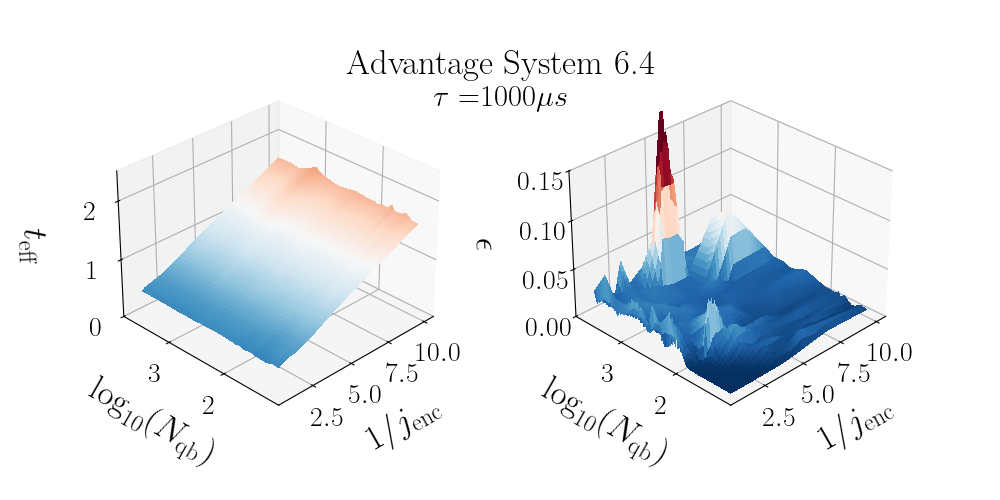}
        \caption{}
    \end{subfigure}  \\
    \begin{subfigure}[b]{0.48\textwidth}
        \centering
        \includegraphics[width=\textwidth]{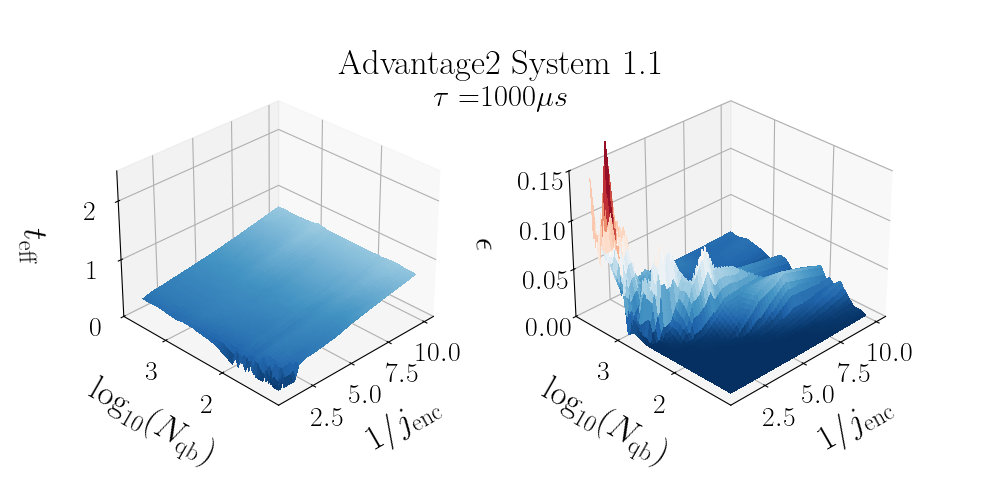}
        \caption{}
    \end{subfigure}
    \quad
    \begin{subfigure}[b]{0.48\textwidth}
        \centering
        \includegraphics[width=\textwidth]{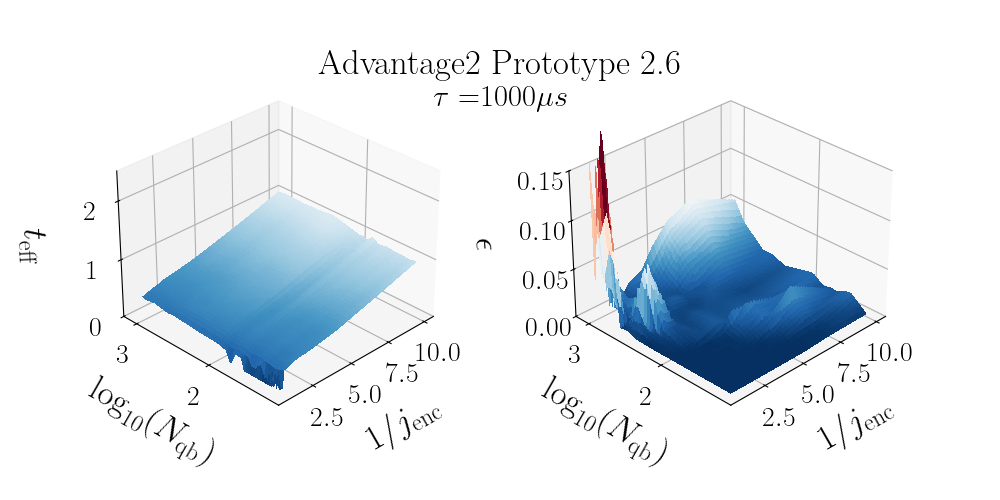}
        \caption{}
    \end{subfigure}
    \caption{\textbf{Effective Temperatures and  TVD Deviations from Gibbs Sampling} for the four machines at annealing time of $\tau= 1000 \mu s$. Each sub-figure corresponds to a different machine and contains 3D Surface plots of the effective temperature ($\Te$), and TVD ($\epsilon$) plotted against the inverse encoded energy coupling $1/\Jen$ and order of magnitude of system size $\log_{10}(\Nqb)$. The small gold colored regions in the plots for $\epsilon$ for Advantage2 devices correspond to zero temperature samples, for which $\epsilon$ is not plotted: as explained in the text, strong energy couplings and small system sizes lead to inapplicable sampling distributions and a breakdown of our thermometry methods, because the ground state is always reached.}
    \label{fig:surface_plot_S1}
\end{figure*}

\begin{figure*}[h!!!]

     \centering
    \begin{subfigure}[b]{0.48\textwidth}
        \centering
        \includegraphics[width=\textwidth]{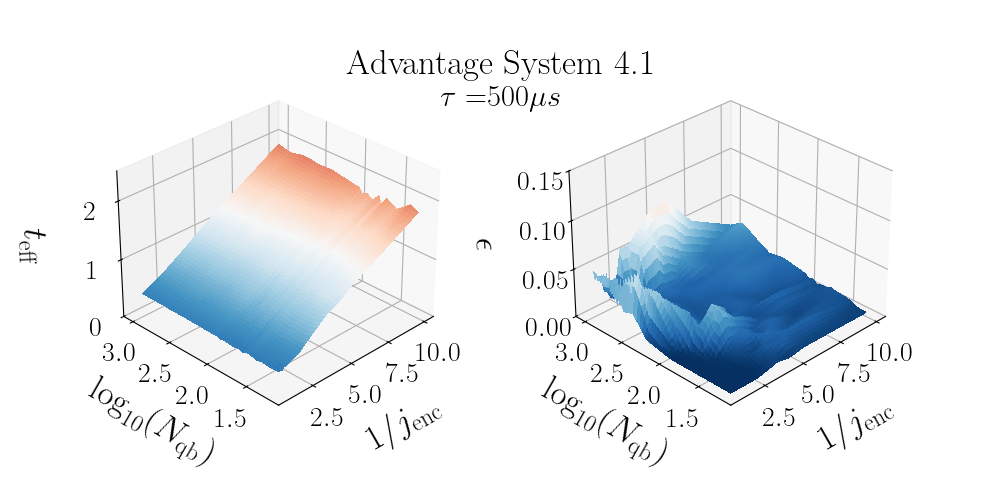}
        \caption{}
    \end{subfigure}
    \quad
    \begin{subfigure}[b]{0.48\textwidth}
        \centering
        \includegraphics[width=\textwidth]{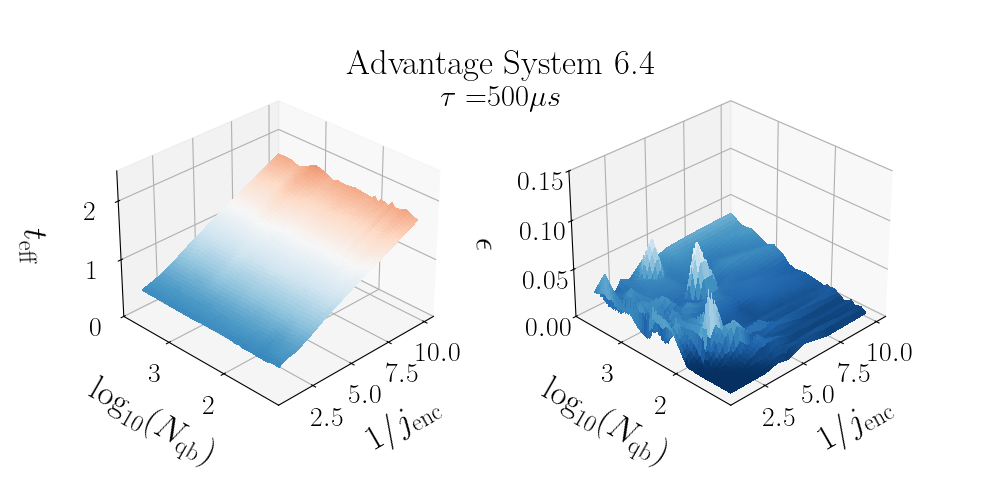}
        \caption{}
    \end{subfigure}  \\
    \begin{subfigure}[b]{0.48\textwidth}
        \centering
        \includegraphics[width=\textwidth]{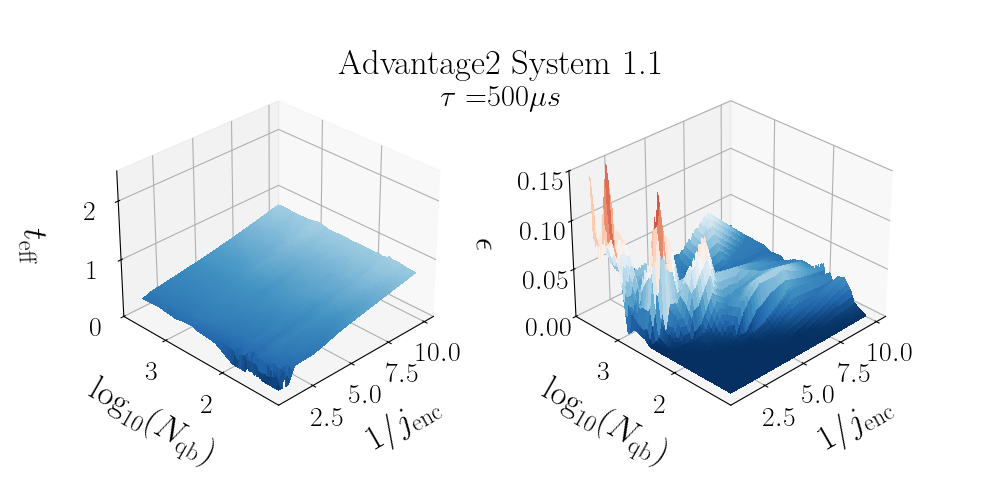}
        \caption{}
    \end{subfigure}
    \quad
    \begin{subfigure}[b]{0.48\textwidth}
        \centering
        \includegraphics[width=\textwidth]{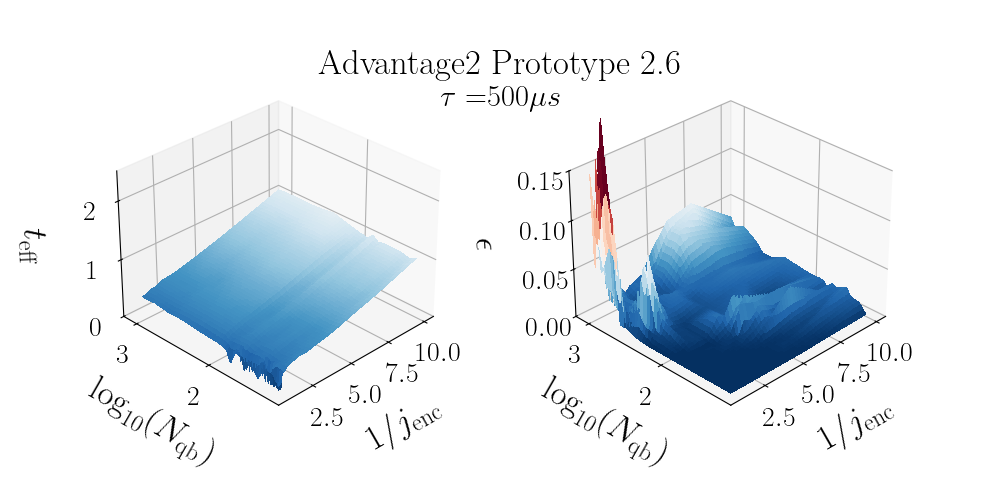}
        \caption{}
    \end{subfigure}
    \caption{\textbf{Effective Temperatures and  TVD Deviations from Gibbs Sampling} for the four machines at annealing time of $\tau= 500 \mu s$. Each subfigure corresponds to a different machine and contains 3D Surface plots of the effective temperature ($\Te$), and TVD ($\epsilon$) plotted against the inverse encoded energy coupling $1/\Jen$ and order of magnitude of system size $\log_{10}(\Nqb)$. The small gold colored regions in the plots for $\epsilon$ for Advantage2 devices correspond to zero temperature samples, for which $\epsilon$ is not plotted: as explained in the text, strong energy couplings and small system sizes lead to inapplicable sampling distributions and a breakdown of our thermometry methods, because the ground state is always reached.}
    \label{fig:surface_plot_S2}
\end{figure*}

\begin{figure*}[h!!!]

     \centering
    \begin{subfigure}[b]{0.48\textwidth}
        \centering
        \includegraphics[width=\textwidth]{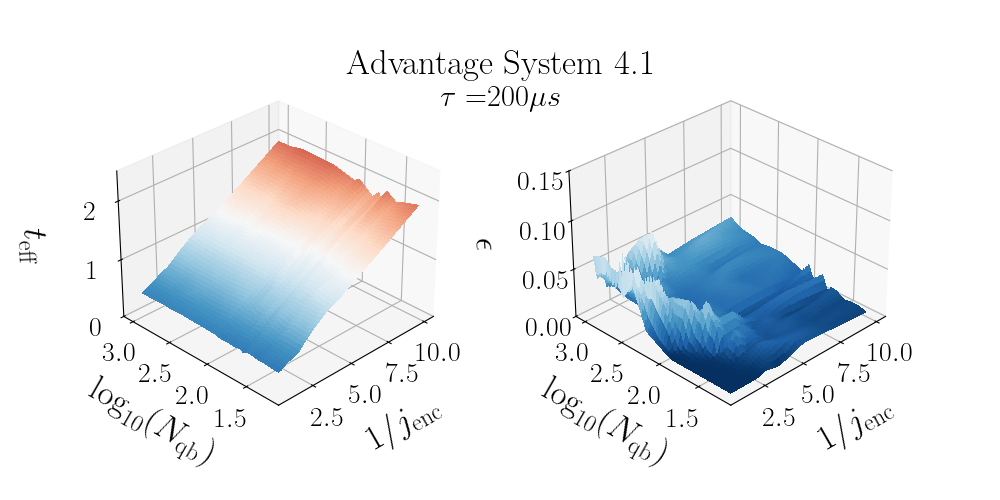}
        \caption{}
    \end{subfigure}
    \quad
    \begin{subfigure}[b]{0.48\textwidth}
        \centering
        \includegraphics[width=\textwidth]{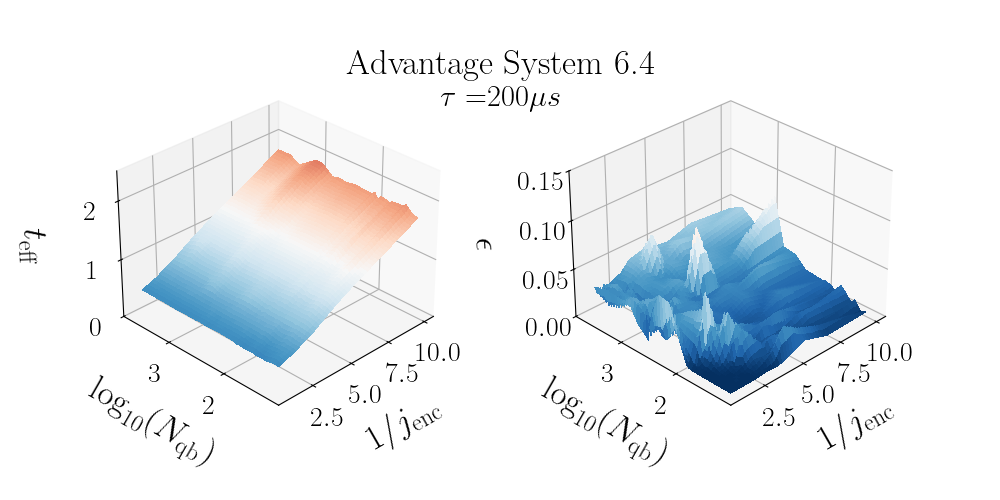}
        \caption{}
    \end{subfigure}  \\
    \begin{subfigure}[b]{0.48\textwidth}
        \centering
        \includegraphics[width=\textwidth]{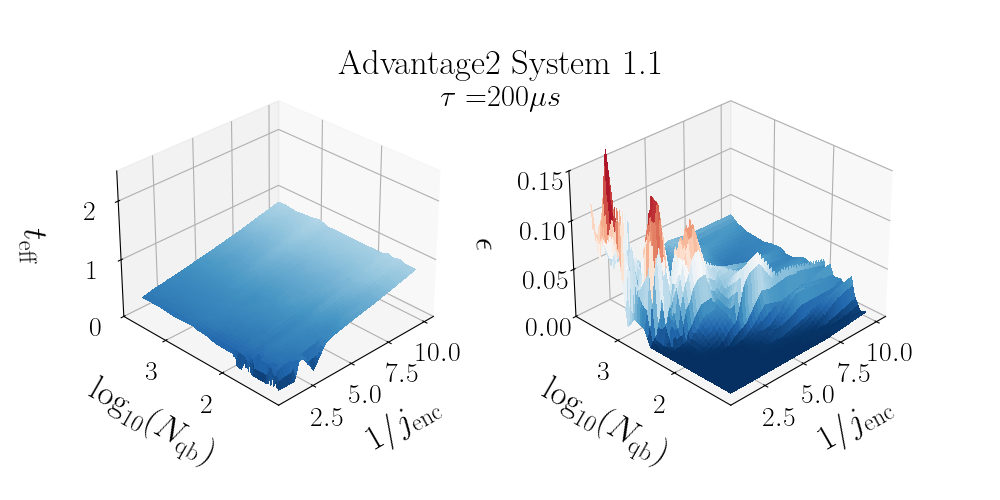}
        \caption{}
    \end{subfigure}
    \quad
    \begin{subfigure}[b]{0.48\textwidth}
        \centering
        \includegraphics[width=\textwidth]{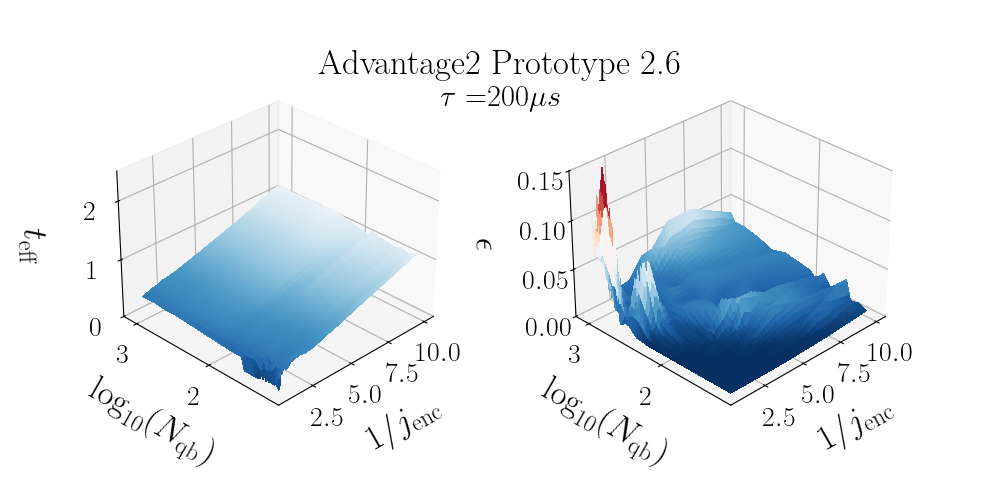}
        \caption{}
    \end{subfigure}
    \caption{\textbf{Effective Temperatures and  TVD Deviations from Gibbs Sampling} for the four machines at annealing time of $\tau= 200 \mu s$. Each sub-figure corresponds to a different machine and contains 3D Surface plots of the effective temperature ($\Te$), and TVD ($\epsilon$) plotted against the inverse encoded energy coupling $1/\Jen$ and order of magnitude of system size $\log_{10}(\Nqb)$. The small gold colored regions in the plots for $\epsilon$ for Advantage2 devices correspond to zero temperature samples, for which $\epsilon$ is not plotted: as explained in the text, strong energy couplings and small system sizes lead to inapplicable sampling distributions and a breakdown of our thermometry methods, because the ground state is always reached.}
    \label{fig:surface_plot_S3}
\end{figure*}

\begin{figure*}[h!!!]

     \centering
    \begin{subfigure}[b]{0.48\textwidth}
        \centering
        \includegraphics[width=\textwidth]{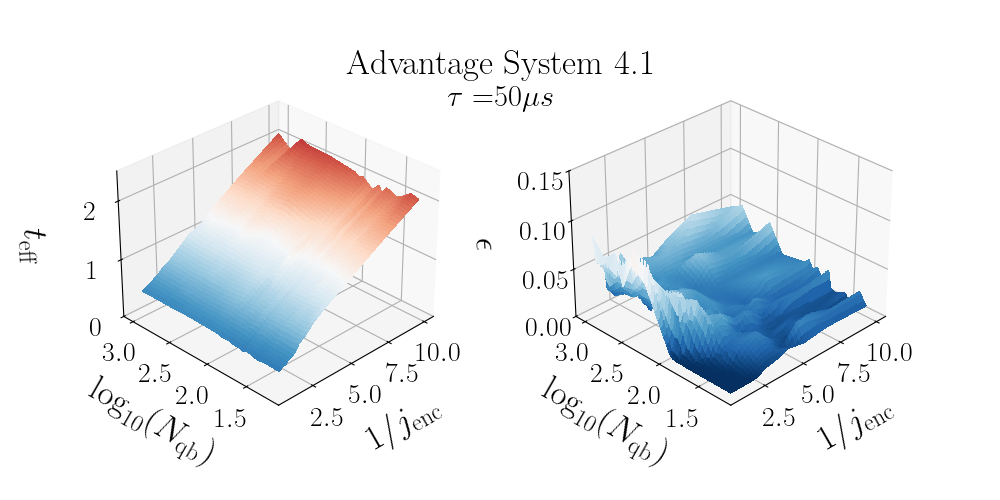}
        \caption{}
    \end{subfigure}
    \quad
    \begin{subfigure}[b]{0.48\textwidth}
        \centering
        \includegraphics[width=\textwidth]{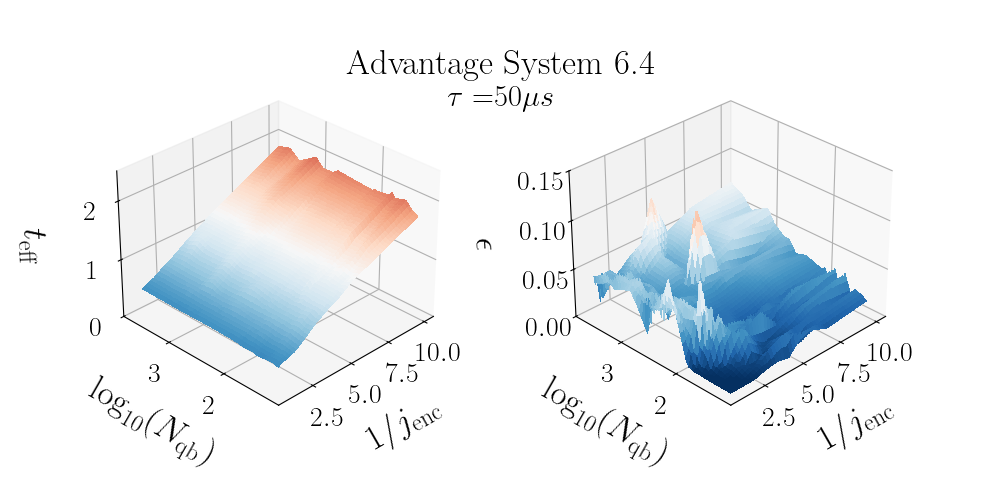}
        \caption{}
    \end{subfigure}  \\
    \begin{subfigure}[b]{0.48\textwidth}
        \centering
        \includegraphics[width=\textwidth]{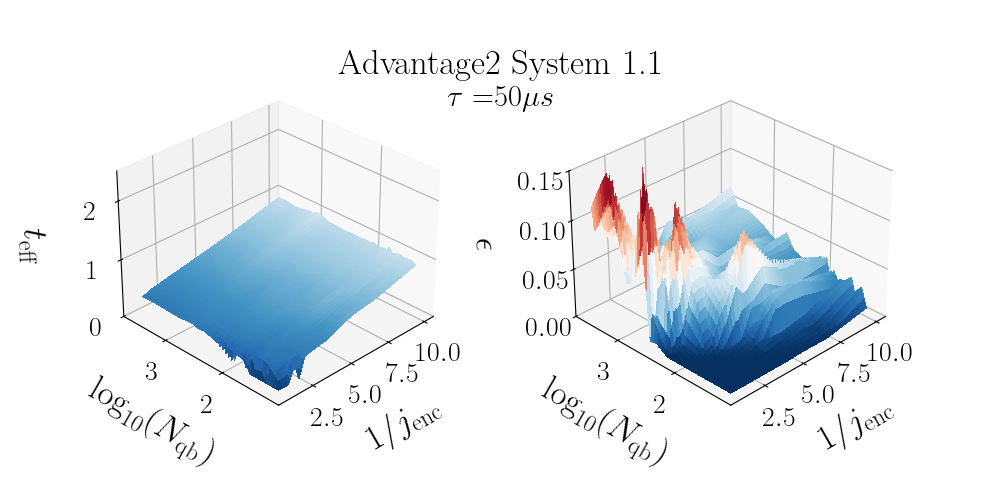}
        \caption{}
    \end{subfigure}
    \quad
    \begin{subfigure}[b]{0.48\textwidth}
        \centering
        \includegraphics[width=\textwidth]{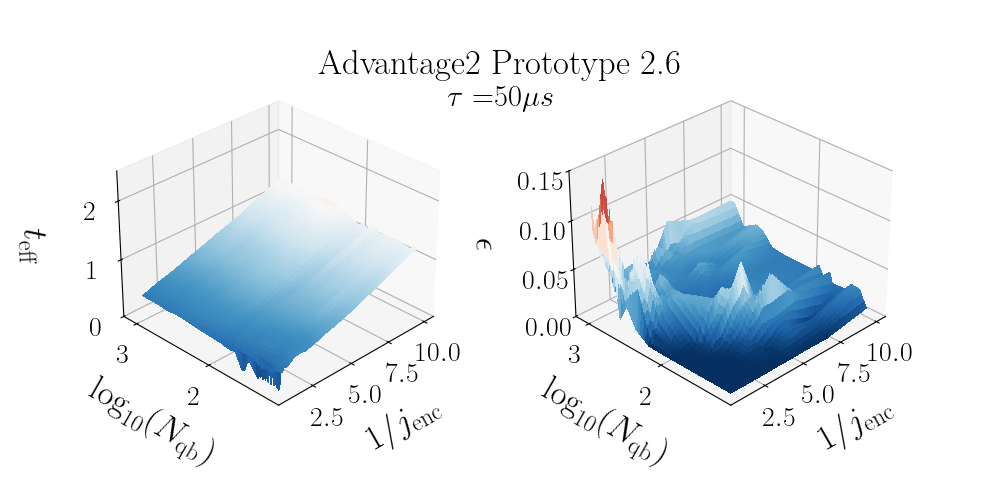}
        \caption{}
    \end{subfigure}
    \caption{\textbf{Effective Temperatures and  TVD Deviations from Gibbs Sampling} for the four machines at annealing time of $\tau= 50 \mu s$. Each subfigure corresponds to a different machine and contains 3D Surface plots of the effective temperature ($\Te$), and TVD ($\epsilon$) plotted against the inverse encoded energy coupling $1/\Jen$ and order of magnitude of system size $\log_{10}(\Nqb)$. The small gold colored regions in the plots for $\epsilon$ for Advantage2 devices correspond to zero temperature samples, for which $\epsilon$ is not plotted: as explained in the text, strong energy couplings and small system sizes lead to inapplicable sampling distributions and a breakdown of our thermometry methods, because the ground state is always reached.} 
    \label{fig:surface_plot_S4}
\end{figure*}
\begin{figure*}[h!!!]

     \centering
    \begin{subfigure}[b]{0.48\textwidth}
        \centering
        \includegraphics[width=\textwidth]{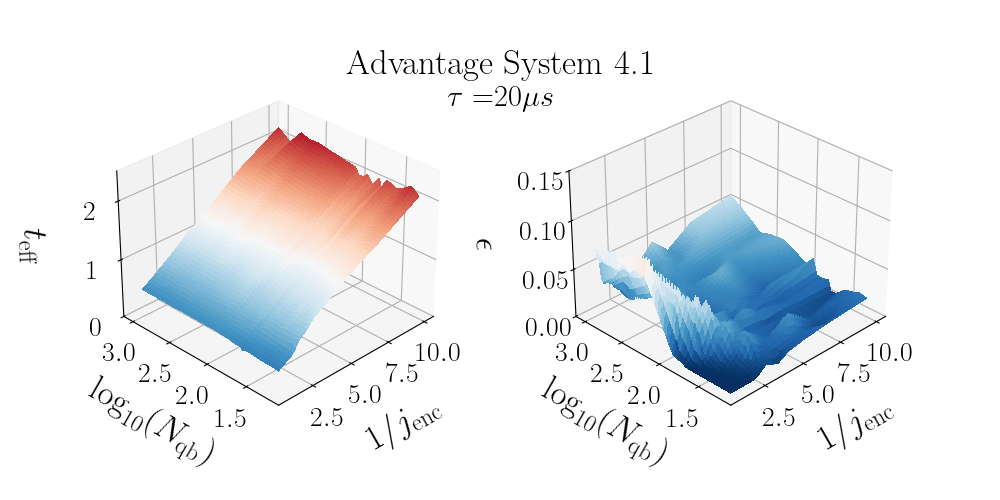}
        \caption{}
    \end{subfigure}
    \quad
    \begin{subfigure}[b]{0.48\textwidth}
        \centering
        \includegraphics[width=\textwidth]{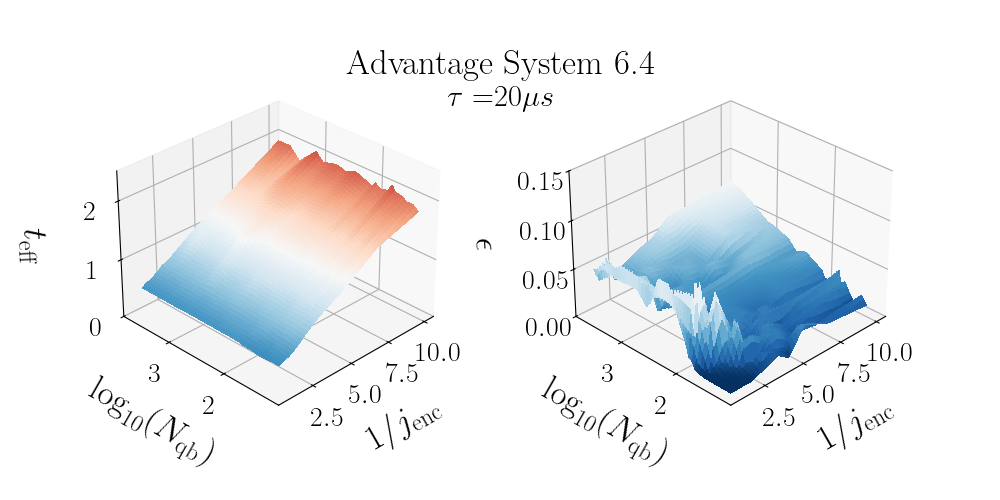}
        \caption{}
    \end{subfigure}  \\
    \begin{subfigure}[b]{0.48\textwidth}
        \centering
        \includegraphics[width=\textwidth]{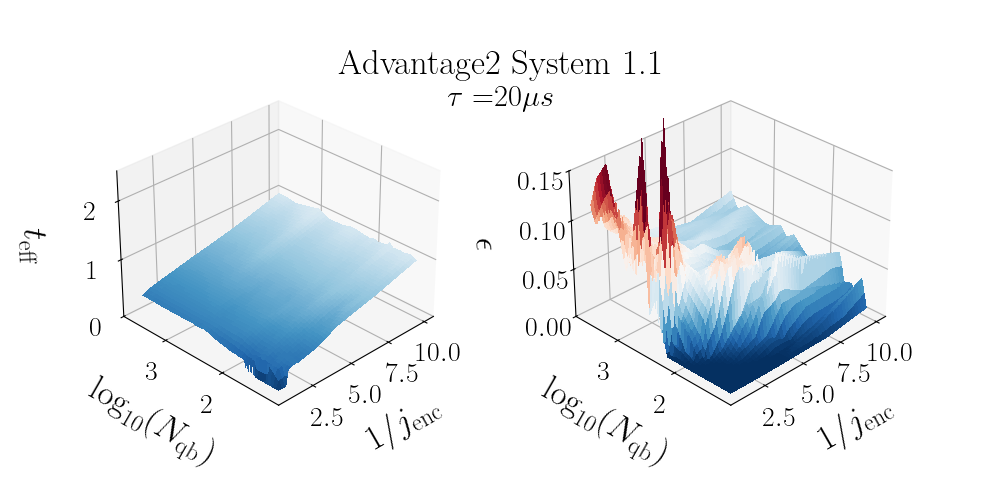}
        \caption{}
    \end{subfigure}
    \quad
    \begin{subfigure}[b]{0.48\textwidth}
        \centering
        \includegraphics[width=\textwidth]{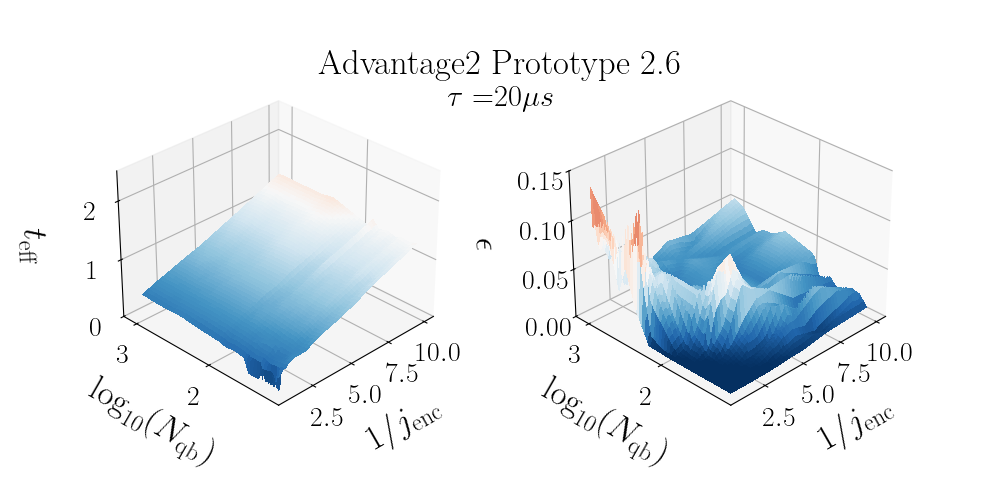}
        \caption{}
    \end{subfigure}
    \caption{\textbf{Effective Temperatures and  TVD Deviations from Gibbs Sampling} for the four machines at annealing time of $\tau= 20 \mu s$. Each sub-figure corresponds to a different machine and contains 3D Surface plots of the effective temperature ($\Te$), and TVD ($\epsilon$) plotted against the inverse encoded energy coupling $1/\Jen$ and order of magnitude of system size $\log_{10}(\Nqb)$. The small gold colored regions in the plots for $\epsilon$ for Advantage2 devices correspond to zero temperature samples, for which $\epsilon$ is not plotted: as explained in the text, strong energy couplings and small system sizes lead to inapplicable sampling distributions and a breakdown of our thermometry methods, because the ground state is always reached. }
    \label{fig:surface_plot_S5}
\end{figure*}
\begin{figure*}[h!!!]
     \centering
    \begin{subfigure}[b]{0.48\textwidth}
        \centering
        \includegraphics[width=\textwidth]{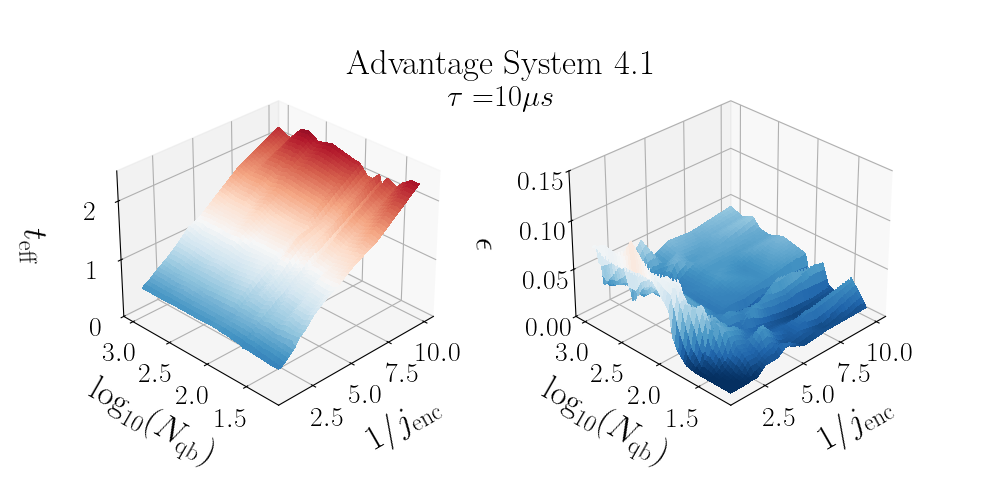}
        \caption{}
    \end{subfigure}
    \quad
    \begin{subfigure}[b]{0.48\textwidth}
        \centering
        \includegraphics[width=\textwidth]{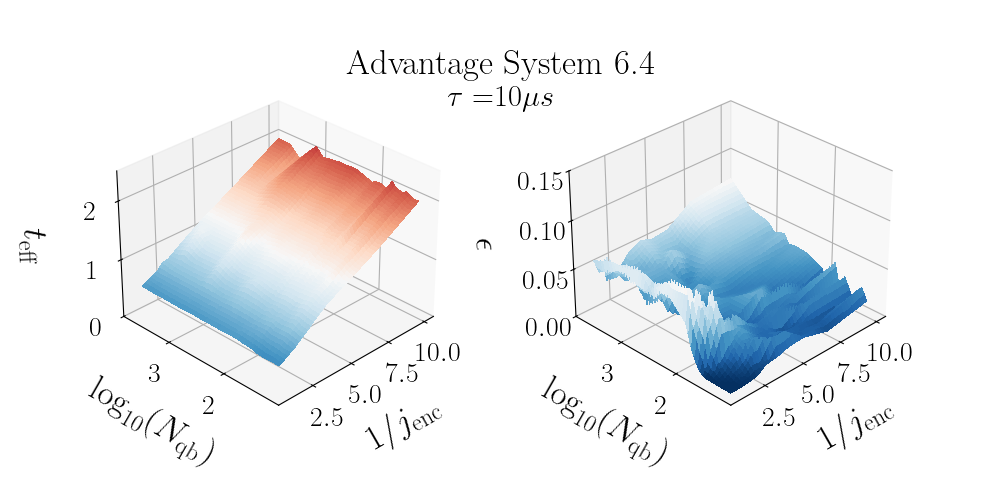}
        \caption{}
    \end{subfigure}  \\
    \begin{subfigure}[b]{0.48\textwidth}
        \centering
        \includegraphics[width=\textwidth]{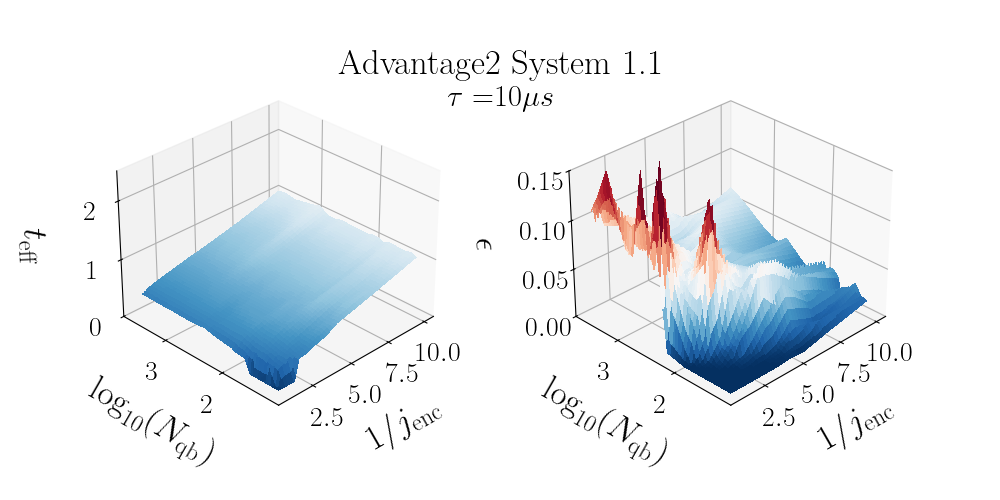}
        \caption{}
    \end{subfigure}
    \quad
    \begin{subfigure}[b]{0.48\textwidth}
        \centering
        \includegraphics[width=\textwidth]{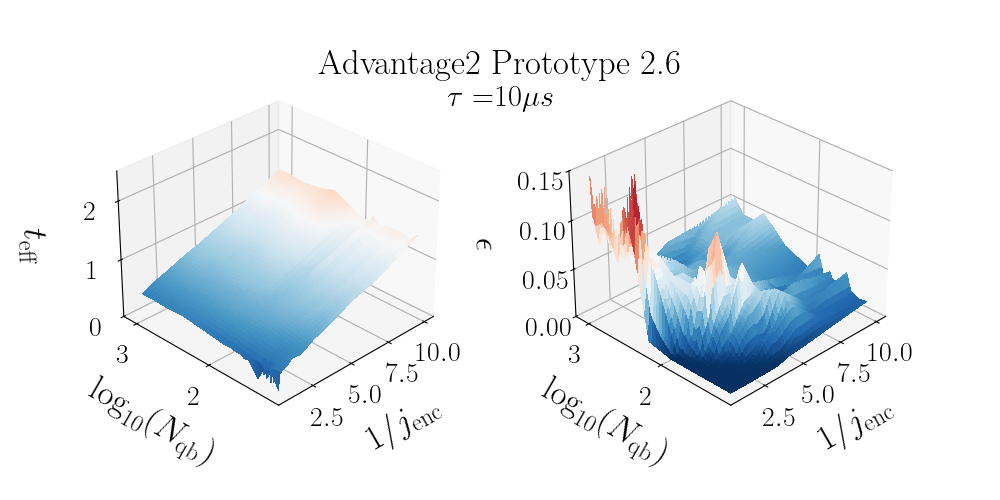}
        \caption{}
    \end{subfigure}
    \caption{\textbf{Effective Temperatures and  TVD Deviations from Gibbs Sampling} for the four machines at annealing time of $\tau= 10 \mu s$. Each sub-figure corresponds to a different machine and contains 3D Surface plots of the effective temperature ($\Te$), and TVD ($\epsilon$) plotted against the inverse encoded energy coupling $1/\Jen$ and order of magnitude of system size $\log_{10}(\Nqb)$. The small gold colored regions in the plots for $\epsilon$ for Advantage2 devices correspond to zero temperature samples, for which $\epsilon$ is not plotted: as explained in the text, strong energy couplings and small system sizes lead to inapplicable sampling distributions and a breakdown of our thermometry methods, because the ground state is always reached.}
    \label{fig:surface_plot_S6}
\end{figure*}

\begin{figure*}[h!!!]
    \centering
    \begin{subfigure}[b]{0.48\textwidth}
        \centering
        \includegraphics[width=\textwidth]{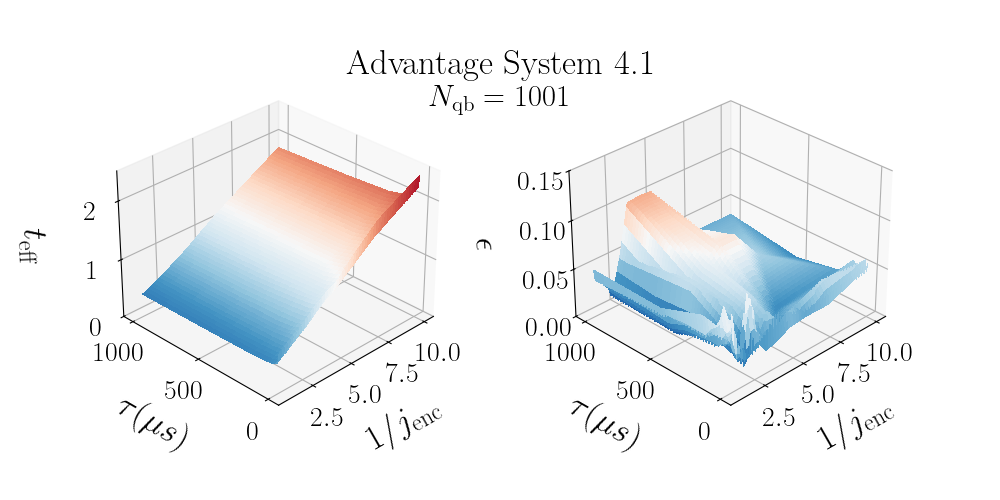}
        \caption{}
    \end{subfigure}
    \quad
    \begin{subfigure}[b]{0.48\textwidth}
        \centering
        \includegraphics[width=\textwidth]{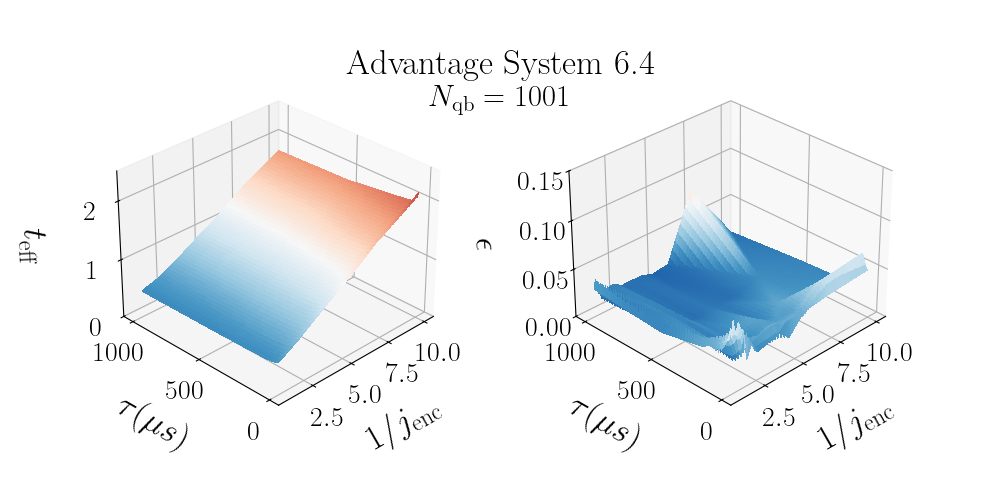}
        \caption{}
    \end{subfigure}  \\
    \begin{subfigure}[b]{0.48\textwidth}
        \centering
        \includegraphics[width=\textwidth]{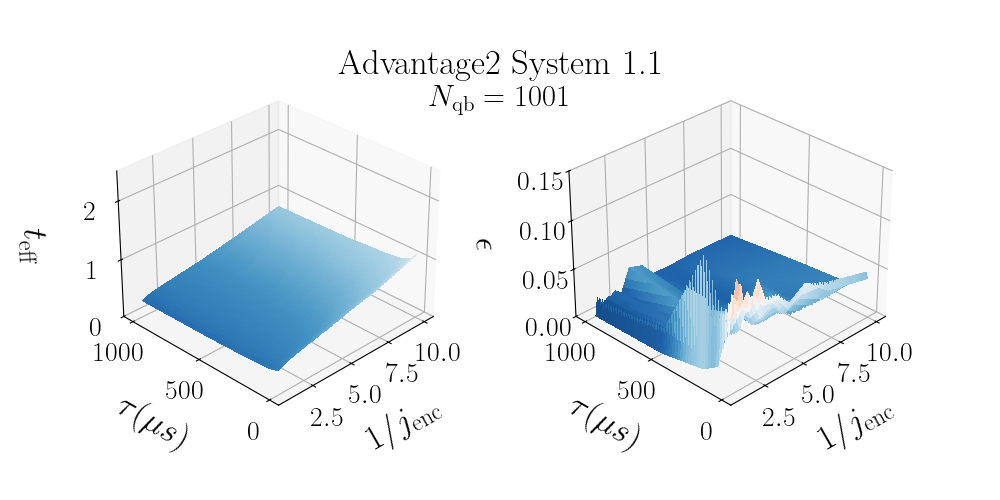}
        \caption{}
    \end{subfigure}
    \quad
    \begin{subfigure}[b]{0.48\textwidth}
        \centering
        \includegraphics[width=\textwidth]{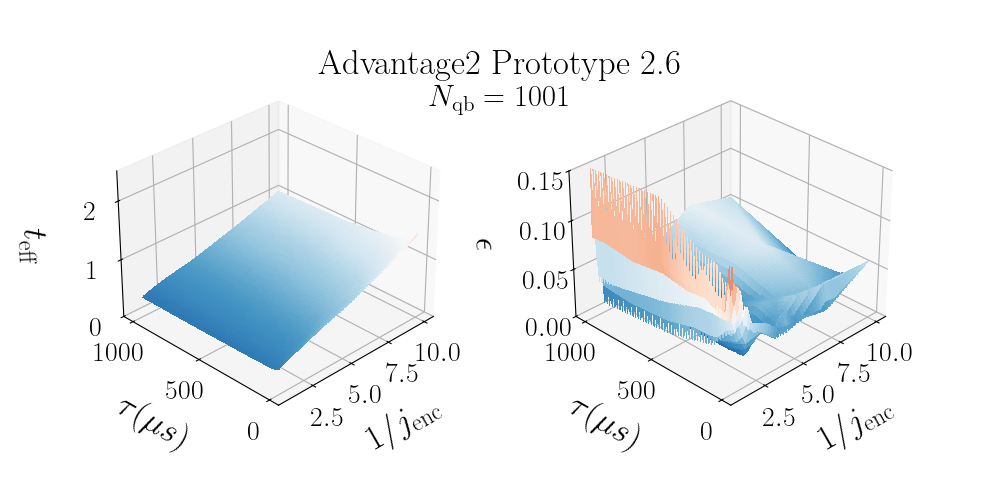}
        \caption{}
    \end{subfigure}
    \caption{{\bf Effective Temperatures and  TVD Deviations from Gibbs Sampling} for the four machines at large size $\Nqb= 1001$. Each sub-figure corresponds to a machine and contains  3D Surface plots of the effective temperature ($\Te$), and TVD ($\epsilon$) plotted versus the inverse encoded energy coupling $1/\Jen$ and annealing time $\tau$. The small gold colored regions in the plots for $\epsilon$ for Advantage2 devices correspond to zero temperature samples, for which $\epsilon$ is not plotted: as explained in the text, strong energy couplings and small system sizes lead to inapplicable sampling distributions and a breakdown of our thermometry methods, because the ground state is always reached.}
    \label{surface_plot_S7}
\end{figure*}

\end{widetext}

\end{document}